\documentclass[fleqn,usenatbib]{mnras}

\usepackage[T1]{fontenc}

\DeclareRobustCommand{\VAN}[3]{#2}
\let\VANthebibliography\thebibliography
\def\thebibliography{\DeclareRobustCommand{\VAN}[3]{##3}\VANthebibliography}

\usepackage{graphicx}
\usepackage{amsmath}
\usepackage{amssymb}
\usepackage{upgreek}
\usepackage{anyfontsize}
\usepackage{pifont}
\usepackage{placeins}
\usepackage{xcolor}
\usepackage{orcidlink}
\usepackage{float}
\usepackage{tikz}
\usetikzlibrary{matrix}

\newcommand{\spectralline}[3]{\ion{#1}{#2} $\uplambda#3$ \AA}
\newcommand{\shortspecline}[2]{\ion{#1}{#2}}
\newcommand{\fspectralline}[3]{[\ion{#1}{#2}] $\uplambda#3$ \AA}
\newcommand{\shortfspecline}[2]{[\ion{#1}{#2}]}
\newcommand{\Ha}{H$\upalpha$\ }

\newcommand{\msun}{\mathrm{M}_\odot}

\newcommand{\mycomment}[1]{}
\newcommand{\change}[1]{#1}
\newcommand{\changetwo}[1]{#1}
\newcommand{\changethree}[1]{#1}
\newcommand{\cmark}{\ding{51}}
\newcommand{\xmark}{\ding{55}}

\newcommand{\galrate}{$R_\mathrm{G}=3.6~^{+2.6}_{-1.8}~(\mathrm{statistical})~^{+5.1}_{-0.0}~(\mathrm{systematic})\times10^{-6}~\mathrm{galaxy}^{-1}~\mathrm{yr}^{-1}$}
\newcommand{\massrate}{$R_\mathrm{M}=3.1~^{+2.3}_{-1.5}~(\mathrm{statistical})~^{+4.4}_{-0.0}~(\mathrm{systematic})\times10^{-17}~\mathrm{M_\odot^{-1}}~\mathrm{yr}^{-1}$}
\newcommand{\volrate}{$R_\mathrm{V}=7~^{+20}_{-5}~(\mathrm{statistical})~^{+10}_{-0.0}~(\mathrm{systematic})\times10^{-9}~\mathrm{Mpc}^{-3}~\mathrm{yr}^{-1}$}

\newcommand{\figref}[1]{Fig. \ref{#1}}
\newcommand{\tabref}[1]{Table \ref{#1}}
\newcommand{\secref}[1]{Section \ref{#1}}
\newcommand{\equref}[1]{equation (\ref{#1})}

\newenvironment{Contfigure*}{%
\renewcommand{\thefigure}{\arabic{figure} (Cont.)}%
\addtocounter{figure}{-1}%
\begin{figure*}}{%
\renewcommand{\thefigure}{\arabic{figure}}%
\end{figure*}}

\title[ECLEs in SDSS]{The rate of extreme coronal line emitting galaxies in the Sloan Digital Sky Survey and their relation to tidal disruption events}

\author[J. Callow et al.]{
\parbox{\textwidth}{
J.~Callow$^{\orcidlink{0000-0002-0804-9533}}$,$^{1}$\thanks{E-mail: joe.callow@port.ac.uk}
O.~Graur$^{\orcidlink{0000-0002-4391-6137}}$,$^{1,2}$
P.~Clark$^{\orcidlink{0000-0002-6576-7400}}$,$^{1}$
A.~Palmese$^{\orcidlink{0000-0002-6011-0530}}$,$^{3}$
J.~Aguilar$^{\orcidlink{0000-0003-0822-452X}}$,$^{4}$
S.~Ahlen$^{\orcidlink{0000-0001-6098-7247}}$,$^{5}$
S.~BenZvi$^{\orcidlink{0000-0001-5537-4710}}$,$^{6}$
D.~Brooks$^{\orcidlink{0000-0002-8458-5047}}$,$^{7}$
T.~Claybaugh$^{\orcidlink{0000-0002-5024-6987}}$,$^{4}$
A.~de la Macorra$^{\orcidlink{0000-0002-1769-1640}}$,$^{8}$
P.~Doel$^{\orcidlink{0000-0002-6397-4457}}$,$^{7}$
J.~E.~Forero-Romero$^{\orcidlink{0000-0002-2890-3725}}$,$^{9,10}$
E.~Gaztañaga$^{\orcidlink{0000-0001-9632-0815}}$,$^{11,1,12}$
S.~Gontcho A Gontcho$^{\orcidlink{0000-0003-3142-233X}}$,$^{4}$
A.~Lambert$^{\orcidlink{0009-0003-5658-2601}}$,$^{4}$
M.~Landriau$^{\orcidlink{0000-0003-1838-8528}}$,$^{4}$
M.~Manera$^{\orcidlink{0000-0003-4962-8934}}$,$^{13,14}$
A.~Meisner$^{\orcidlink{0000-0002-1125-7384}}$,$^{15}$
R.~Miquel$^{\orcidlink{0000-0002-6610-4836}}$,$^{16,14}$
J.~Moustakas$^{\orcidlink{0000-0002-2733-4559}}$,$^{17}$
J.~Nie$^{\orcidlink{0000-0001-6590-8122}}$,$^{18}$
C.~Poppett$^{\orcidlink{0000-0003-0512-5489}}$,$^{4,19,20}$
F.~Prada$^{\orcidlink{0000-0001-7145-8674}}$,$^{21}$
M.~Rezaie$^{\orcidlink{0000-0001-5589-7116}}$,$^{22}$
G.~Rossi,$^{23}$
E.~Sanchez$^{\orcidlink{0000-0002-9646-8198}}$,$^{24}$
J.~Silber$^{\orcidlink{0000-0002-3461-0320}}$,$^{4}$
G.~Tarl\'{e}$^{\orcidlink{0000-0003-1704-0781}}$,$^{25}$
B.~A.~Weaver,$^{15}$
and Z.~Zhou$^{\orcidlink{0000-0002-4135-0977}}$$^{18}$
}
\vspace{0.4cm}
\\
\parbox{\textwidth}{
$^{1}$ Institute of Cosmology \& Gravitation, University of Portsmouth, Dennis Sciama Building, Portsmouth, PO1 3FX, UK\\
$^{2}$ Department of Astrophysics, American Museum of Natural History, New York, NY 10024, USA\\
$^{3}$ Department of Physics, Carnegie Mellon University, 5000 Forbes Avenue, Pittsburgh, PA 15213, USA\\
$^{4}$ Lawrence Berkeley National Laboratory, 1 Cyclotron Road, Berkeley, CA 94720, USA\\
$^{5}$ Physics Dept., Boston University, 590 Commonwealth Avenue, Boston, MA 02215, USA\\
$^{6}$ Department of Physics \& Astronomy, University of Rochester, 206 Bausch and Lomb Hall, P.O. Box 270171, Rochester, NY 14627-0171, USA\\
$^{7}$ Department of Physics \& Astronomy, University College London, Gower Street, London, WC1E 6BT, UK\\
$^{8}$ Instituto de F\'{\i}sica, Universidad Nacional Aut\'{o}noma de M\'{e}xico,  Cd. de M\'{e}xico  C.P. 04510,  M\'{e}xico\\
$^{9}$ Departamento de F\'isica, Universidad de los Andes, Cra. 1 No. 18A-10, Edificio Ip, CP 111711, Bogot\'a, Colombia\\
$^{10}$ Observatorio Astron\'omico, Universidad de los Andes, Cra. 1 No. 18A-10, Edificio H, CP 111711 Bogot\'a, Colombia\\
$^{11}$ Institut d'Estudis Espacials de Catalunya (IEEC), 08034 Barcelona, Spain\\
$^{12}$ Institute of Space Sciences, ICE-CSIC, Campus UAB, Carrer de Can Magrans s/n, 08913 Bellaterra, Barcelona, Spain\\
$^{13}$ Departament de F\'{i}sica, Serra H\'{u}nter, Universitat Aut\`{o}noma de Barcelona, 08193 Bellaterra (Barcelona), Spain\\
$^{14}$ Institut de F\'{i}sica d’Altes Energies (IFAE), The Barcelona Institute of Science and Technology, Campus UAB, 08193 Bellaterra Barcelona, Spain\\
$^{15}$ NSF's NOIRLab, 950 N. Cherry Ave., Tucson, AZ 85719, USA\\
$^{16}$ Instituci\'{o} Catalana de Recerca i Estudis Avan\c{c}ats, Passeig de Llu\'{\i}s Companys, 23, 08010 Barcelona, Spain\\
$^{17}$ Department of Physics and Astronomy, Siena College, 515 Loudon Road, Loudonville, NY 12211, USA\\
$^{18}$ National Astronomical Observatories, Chinese Academy of Sciences, A20 Datun Rd., Chaoyang District, Beijing, 100012, P.R. China\\
$^{19}$ Space Sciences Laboratory, University of California, Berkeley, 7 Gauss Way, Berkeley, CA  94720, USA\\
$^{20}$ University of California, Berkeley, 110 Sproul Hall \#5800 Berkeley, CA 94720, USA\\
$^{21}$ Instituto de Astrof\'{i}sica de Andaluc\'{i}a (CSIC), Glorieta de la Astronom\'{i}a, s/n, E-18008 Granada, Spain\\
$^{22}$ Department of Physics, Kansas State University, 116 Cardwell Hall, Manhattan, KS 66506, USA\\
$^{23}$ Department of Physics and Astronomy, Sejong University, Seoul, 143-747, Korea\\
$^{24}$ CIEMAT, Avenida Complutense 40, E-28040 Madrid, Spain\\
$^{25}$ University of Michigan, Ann Arbor, MI 48109, USA\\
}
}

\date{Accepted XXX. Received YYY; in original form ZZZ}

\pubyear{2023}

\usepackage{newtxtext,newtxmath}

\label{firstpage}
\pagerange{\pageref{firstpage}--\pageref{lastpage}}
\begin{document}
\maketitle

\defcitealias{Wang2012}{W12}
\defcitealias{Clark2024}{Clark et al. 2024}

\begin{abstract}
High-ionization iron coronal lines (CLs) are a rare phenomenon observed in galaxy and quasi-stellar object spectra that are thought to be created \change{by high-energy emission from active galactic nuclei and certain types of transients.
In cases known as extreme coronal line emitting galaxies (ECLEs), these CLs are strong and fade away on a timescale of years.
The most likely progenitors of these variable CLs are tidal disruption events (TDEs), which produce sufficient high-energy emission to create and sustain the CLs over these timescales.
To test the possible connection between ECLEs and TDEs, we present the most complete variable ECLE rate calculation to date and compare the results to TDE rates from the literature.
To achieve this, we search for ECLEs in the Sloan Digital Sky Survey (SDSS).}
We detect sufficiently strong CLs in \change{16} galaxies, more than doubling the number previously found in SDSS.
Using follow-up spectra from the Dark Energy Spectroscopic Instrument and Gemini Multi-Object Spectrograph, \textit{Wide-field Infrared Survey Explorer} mid-infrared observations, and Liverpool Telescope optical photometry, \change{we find that none of the nine new ECLEs evolve in a manner consistent with that of the five previously discovered variable ECLEs.}
\change{Using this sample of five variable ECLEs}, we calculate the galaxy-normalized rate of \change{variable} ECLEs in SDSS to be \galrate.
The mass-normalized rate is \massrate\ and the volumetric rate is \volrate.
\change{Our rates are one to two orders of magnitude lower than TDE rates from the literature, \changetwo{which suggests that only 10 to 40 per cent of all TDEs produce variable ECLEs.}}
\changethree{Additional uncertainties in the rates arising from the structure of the interstellar medium have yet to be included.}
\end{abstract}


\section{Introduction}
\label{sec:intro}

A tidal disruption event occurs when a star's orbit takes it within the tidal radius of a supermassive black hole (SMBH), where the tidal forces of the black hole overcome the star's internal gravity \citep{Hills1975}.
During the star's ensuing disruption, a portion of the star's matter falls onto the black hole and a flare of electromagnetic emission is produced.
The origin of this flare is debated, with options including circularization of the infalling matter to form a temporary accretion disc \citep{Komossa2023} and shocks caused by collisions within the stream \citep{Fancher2023}.
Whichever process or processes do cause the emission, a TDE creates a range of observational signatures across the entire electromagnetic spectrum \citep{vanVelzen2020,Alexander2020,Saxton2021}.

TDEs provide a novel method to study SMBHs at a lower mass range than previously possible.
The most established methods of measuring SMBH masses are modelling the kinematics of orbiting stars or gas (\citealp{Sargent1978,Gebhardt2000,Siopis2009,Roberts2021}; \citealp[see review by][]{Kormendy2013}), and reverberation mapping, which involves measuring the response of the gas in an active galactic nucleus (AGN) to changes in the continuum flux \citep{Blandford1982,Peterson1993,Kaspi2000,Cackett2021}.
However, these methods are most effective for SMBHs with masses $\gtrsim10^6~\mathrm{M}_\odot$.
Lower-mass, quiescent black holes are more difficult to study as they have less of an effect on their surroundings \citep{Greene2020}.
As the latter account for $\sim$90 per cent of the black hole population \citep{Soria2006}, it is important to explore new methods to study them.
TDEs provide such a method as they are only visible when they occur around SMBHs for which the tidal radius is larger than the event horizon.
For a Sun-like star, this corresponds to a maximum black hole mass $M_\mathrm{BH}\approx10^8~\mathrm{M}_\odot$.
Black holes with masses as low as $10^5 -10^6~\mathrm{M}_\odot$ have been measured using TDE emission \citep{Mockler2019,Ryu2020}.
These black holes lie at the low-mass end of the $M_\mathrm{BH}-\sigma$ relation, which is currently underexplored due to a lack of low-mass detections \citep{Greene2020}.

The first TDEs were discovered in the X-ray regime by the \textit{ROSAT} All Sky Survey \citep{Voges1993}.
This initial survey revealed five galaxies that underwent an X-ray outburst \change{\citep{Bade1996,Komossa1999a,Grupe1999,Komossa1999b}} but showed little to no activity in follow-up observations \citep{Donley2002}.
During the following 20 yr, the \textit{Chandra X-ray Observatory} \citep{Weisskopf2000} and \textit{XMM-Newton} \citep{Jansen2001} increased the number of detected X-ray TDE candidates to $\sim$30 \citep{Esquej2007, Maksym2013, Lin2015}.
The observed flares were well described by black bodies with temperatures in the range $kT_\mathrm{bb}=0.04-0.12~\mathrm{keV}$ \citep{Saxton2021}, supporting the model which stated that the X-ray emission was created by the accretion of the bound material \citep{Rees1988}.
This model also hypothesized that the rate at which the bound matter would fall onto the black hole would be proportional to $t^{-\frac{5}{3}}$, and therefore the luminosity of the emission from a TDE should follow the same power law, i.e.,
\begin{equation}
	L\propto t^{-\frac{5}{3}}.
\end{equation}

Though some TDE light curves have been observed to follow this shape \citep{Gezari2021}, other power law indices have been suggested based on different mechanisms that could produce emission in a TDE, with power law indices of $-19/16$ for viscous disc accretion \citep{Cannizzo1990}, $-5/12$ for disk emission \citep{Lodato2011}, and $-4/3$ for advective, super-Eddington, slim disk accretion \citep{Cannizzo2009,Cannizzo2011}.
\change{It has also been predicted that the power law indices can be different depending on the spectral band observed \citep{Lodato2011}.
This work showed that although the X-ray and bolometric luminosities follow $t^{-\frac{5}{3}}$ for a significant portion of their evolution, the optical and ultraviolet (UV) bands would decay much slower, as $t^{-\frac{5}{12}}$.}
Observational studies have also found that the range of power law indices is much larger than expected.
\citet{Auchettl2017} found that in a sample of 13 X-ray TDEs, the power law indices of the light curves ranged from  $-0.26\pm0.10$ to $-1.89\pm0.20$.
This discrepancy between the theoretically predicted and observationally derived power law indices is still not fully understood.

Currently, the number of detected TDEs stands at roughly 150.
Despite first being detected in the X-ray regime, in recent years most TDEs have been discovered in the optical and UV.
These TDEs were discovered in a range of surveys, such as the \textit{Galaxy Evolution Explorer} \citep{Martin2005,Gezari2006,Gezari2008}, Sloan Digital Sky Survey \citep[SDSS;][]{York2000,vanVelzen2011}, All-Sky Automated Survey for Supernovae \citep[ASASSN;][]{Shappee2014,Holoien2014,Short2020} and Zwicky Transient Facility \citep[ZTF;][]{Bellm2019,vanVelzen2021a,Hammerstein2023}.
Despite their growing number, the origin of the emission in optically/UV detected TDEs is still unclear.
It is thought that the optical/UV emission is produced by a different mechanism (or mechanisms) to the X-ray emission \citep{Ulmer1997}.
The current main theories for the optical/UV origin of this emission are matter stream collisions, where the infalling matter stream collides with itself as it circularizes \citep{Piran2015} or reprocessing of high-energy photons, which results from the X-ray radiation being absorbed by outflowing material, then re-emitted at longer wavelengths \citep{Guillochon2013,Roth2016}.
Alternatively, the emission could also be the result of a combination of these factors \citep{Jiang2016,Lu2020}.

As well as X-ray and optical/UV emission, TDEs have been observed emitting radio waves \citep{Alexander2020}, high-energy neutrinos \citep{Stein2021}, and gamma rays \citep{Colle2020}, and producing infrared (IR) echoes \citep{vanVelzen2021b}.
Though these observations are rarer than X-ray and optical/UV detections, they provide further evidence for the range of emission mechanisms possible in TDEs.

The rate at which TDEs occur can aid the understanding of the processes which cause stars to fall onto black holes and the emission processes that result.
Initial work on the theory behind these events centred on the idea of the `loss cone,' the set of angular momenta for which a star orbiting an SMBH would become a TDE \citep{Stone2020}.
Theoretical calculations predict a TDE rate of $\sim10^{-4}\ \mathrm{galaxy}^{-1}\ \mathrm{yr}^{-1}$ \citep{Magorrian1999,Wang2004,Stone2016}.
However, there is a discrepancy between these theoretically predicted rates and those derived from observations.
The theoretically predicted value agrees with some values derived from observational studies \citep{Esquej2008,Maksym2010,vanVelzen2018,Hung2018}, whereas others are at least an order of magnitude lower \citep{Donley2002,Khabibullin2014,vanVelzen2014,Holoien2016}.
The reason for this discrepancy is still unclear.

The mid-infrared (MIR) may provide a new regime in which to detect TDEs.
\citet{Jiang2016} observed a MIR echo in the TDE ASASSN-14li at a lag of $\sim$36 d to the optical detection.
\citet{Jiang2021} constructed a sample of MIR outbursts in nearby galaxies (MIRONGs) using \textit{Wide-field Infrared Survey Explorer (WISE)} light curves.
Only 10 per cent of these MIRONGs had optical counterparts.
They proposed that the majority of these outbursts were dust-induced light echoes of transient accretion onto SMBHs, which were produced by TDEs or changing-state AGN.
\citet{Wang2022a} analysed spectroscopic follow-up of a portion of the sample and concluded that determining the nature of the MIR echo progenitor from spectral variability alone was not feasible.
Analysis of the MIRONG host galaxies by \citet{Dodd2023} called into question the range of proposed progenitors and found that the majority of MIRONGs were consistent with changing-state AGN rather than hidden TDEs.
\change{However, a search of \textit{NEOWISE} data by \citet{Masterson2024} has revealed $\sim20$ TDE candidates, the majority of which have no optical counterpart.
Therefore, though MIR echoes can be created by changing-state AGN and TDEs, it is possible to discern differences in their light curves.}

TDEs could be discovered days to years later by searching for their signatures in the spectra of galaxies and AGN.
TDE spectra exhibit a range of emission lines, some of which are difficult to distinguish from other transient phenomena such as supernovae (SNe) and changing-state AGN \citep{Charalampopoulos2022}.
However, a possible new signature of TDE activity is a set of high-ionization coronal emission lines.
Galaxies that exhibit these lines are known as extreme coronal line emitters (ECLEs).
The coronal lines (CLs) of interest are \fspectralline{Fe}{vii}{3759}, \fspectralline{Fe}{vii}{5160}, \fspectralline{Fe}{vii}{5722}, \fspectralline{Fe}{vii}{6088}, \fspectralline{Fe}{x}{6376}, \fspectralline{Fe}{xi}{7894}, and \fspectralline{Fe}{xiv}{5304} (\citealt{Komossa2008, Wang2011}; hereafter, the higher ionization CLs will be referred to as \shortfspecline{Fe}{x}, \shortfspecline{Fe}{xi}, and \shortfspecline{Fe}{xiv}, respectively).
The CLs are thought to be created by \change{high-energy emission ionizing the interstellar medium (ISM).}
The ionization potentials of the CLs correspond to a strong ionizing continuum that extends from the UV into the X-ray.

The first ECLE to be discovered was SDSS J095209.56+214313.3 (hereafter SDSS J0952; \citealt{Komossa2008}), which was noted as unusual due to the presence of many strong high-ionization CLs in its SDSS spectrum.
Broad Balmer features and \spectralline{He}{ii}{4686} emission provided further evidence of a strong ionizing continuum that created these emission lines.
Follow-up spectra taken two years after the SDSS spectrum revealed that the strengths of the coronal, Balmer, and HeII emission lines had all decreased relative to the \fspectralline{O}{iii}{5007} line \citep{Komossa2009}.
This variability raised the question of what phenomenon was creating these CLs, with initial suggestions being a TDE, \change{changing-state} AGN, or SN.

A second ECLE, SDSS J074820.67+471214.3 (hereafter SDSS J0748), was discovered by \citet{Wang2011}, although its emission lines were slightly different to those in SDSS J0952.
SDSS J0748 did not have any \shortfspecline{Fe}{vii} or \spectralline{He}{ii}{4686} emission lines but did show broad emission features which peaked at around 4050, 4600, and 6560 \AA\  with widths of several hundred \AA ngstroms.
These features had previously been observed in the spectra of Type II-P SNe \citep{Quimby2007}, although CLs as strong as those in SDSS J0748 had never been observed in a SN spectrum.
Follow-up observations taken four years after the SDSS spectrum showed that all the CLs and broad features had disappeared completely while the \fspectralline{O}{iii}{5007} strength had increased by a factor of ten. \citet{Wang2011} concluded that though the broad features were consistent with previous observations of SNe, this could not explain the strong CLs and so a TDE was the most likely cause, with the features originating in the ejected tidal debris \citep{Strubbe2009}.
The broad emission features seen in SDSS J0748 have also been observed in other TDEs and as such it was included in the sample of spectroscopic TDEs by \citet{Arcavi2014}.

A search through SDSS data release 7 (DR7; \citealt{Abazajian2009}) by \citet[hereafter \citetalias{Wang2012}]{Wang2012} discovered five more galaxies that had strong high-ionization CLs.
Follow-up observations by \citetalias{Wang2012} and \citet{Yang2013}, taken between four and ten years after the SDSS spectra, revealed that the CLs were fading or had disappeared in four of the seven ECLEs.
\citet{Yang2013} considered TDEs to be the transient phenomenon most likely to create the variable CLs.
This was largely due to the very high luminosities of the CLs compared to \fspectralline{O}{iii}{5007}.
The seven ECLEs from the \citetalias{Wang2012} sample were reanalysed by \citet{Clark2024} using follow-up spectra taken 15-19 yr after the SDSS spectra.
They confirmed the disappearance of the CLs in the four ECLEs classified as variable by \citet{Yang2013} but found that the CLs had disappeared in one of the non-variable ECLEs.
Therefore, of the seven ECLEs in the \citetalias{Wang2012} sample, five had disappearing CLs and are thought to be associated with TDEs.

\change{Type IIn SNe have been observed to produce long-lasting CLs \citep{Fransson2002,Izotov2009,Smith2009,Stritzinger2012,Fransson2014}, but the luminosities of the lines are typically factors of hundreds to thousands lower than those in ECLEs \citep{Komossa2009,Wang2011}}.
CLs also frequently appear in AGN \citep{Gelbord2009}, but these are also typically weaker than those in ECLEs.
This is most clearly seen when comparing the CL luminosities to \fspectralline{O}{iii}{5007}.
In AGN, the CLs' luminosities are typically only a few per cent of the \shortfspecline{O}{iii} line \citep{Nagao2000}, whereas in the ECLEs, they have comparable luminosities \citepalias{Wang2012}.
The other main distinguishing feature is the timescales on which the CLs vary.
In ECLEs, they have still been detectable months to years after first detection, far longer than would be expected for SNe \citep{Palaversa2016}.
\change{However, CLs observed in SN 2005ip were still observable $\sim3000$ d after they appeared \citep{Smith2017}.
This is much more similar to the variable ECLEs.}
ECLEs vary on longer timescales than AGN and are less erratic in their variation.
AGN typically vary on timescales of weeks to months and the amplitude of their variation is only a few per cent of their overall luminosity \citep{Hawkins2002}.
These differences in variation timescales are consistent when converted to rest frame timescales.

\change{As with TDEs, MIR observations have revealed more about the nature of ECLEs.}
Analysis of the \citetalias{Wang2012} sample's MIR properties revealed that the ECLEs with variable CLs all showed a decline in the MIR since the SDSS spectra were taken, whereas the non-variable ECLEs were roughly constant \citep{Dou2016,Clark2024}.
Furthermore, it was shown that, using the MIR colour criterion from \citet{Stern2012}, the variable ECLEs evolved from an AGN-like state to a non-AGN state, while the non-variable ECLEs remained consistent with AGN, further reinforcing the difference in the progenitors of the CLs \citep{Clark2024,Hinkle2024}.
\citet{Hinkle2024} also compared ECLE, TDE, and MIRONG host galaxy properties and found some overlap in the populations, providing another link between ECLEs and TDEs.
\change{The MIR echo and CLs are thought to be the result of different wavelengths of the transient emission interacting with dust and gas around the SMBH.
The MIR comes from UV emission being reprocessed by the dust, whereas the CLs are produced by extreme UV/X-rays ionizing the gas and then being reprocessed to optical emission.}

Recently, CLs have been observed appearing in the spectra of the optically discovered \change{TDEs AT2017gge \citep{Onori2022,Wang2022b}, AT2018bcb \citep{Neustadt2020}, AT2019qiz \citep{Short2023}, AT2021dms \citep{Hinkle2024}, AT2021qth \citep{Yao2023}, AT2021acak \citep{Li2023}, AT2022fpx \citep{Koljonen2024}, and AT2022upj \citep{Newsome2024}.}
The optical/UV light curves of these transients match the expected shape of a TDE.
\change{Even within this small sample, there is variation in the nature of the X-ray emission and CLs.
Some exhibit CLs developing shortly after an X-ray flare (AT2017gge and AT2018bcb), whereas others develop CLs closer to the optical peak and long before the subsequent X-ray flare (AT2019qiz, AT2022fpx, and AT2022upj).}
These CL-TDEs are the first direct evidence for strong CLs being created by TDEs and show the value of long-duration spectroscopic follow-up of TDEs.

A test of whether variable ECLEs are produced by TDEs is to compare the rates at which they occur.
\change{In order to perform a full rate calculation, we first} repeat the search for ECLEs in the SDSS Legacy Survey performed by \citetalias{Wang2012}, but using the updated DR17 \citep{DR172022} and our own detection algorithm.
We start in \secref{sec:data} by describing the SDSS galaxy and quasi-stellar object (QSO) sample we search over and the follow-up observations we use to further analyse the ECLEs we discover.
In \secref{sec:CL_galaxies}, we examine the results of running our detection algorithm on the galaxy sample and determine the detection efficiency of our algorithm.
In \secref{sec:ecle_sample}, we describe the cuts made to our sample of galaxies exhibiting CLs (hereby CL galaxies) to determine which ones are ECLEs.
We find \change{16 ECLEs, just over} doubling the previous sample from \citetalias{Wang2012}.
To determine the variable nature of these ECLEs, we analyse them individually using follow-up spectra and MIR observations.
We find that the CLs have not disappeared completely in any of the new ECLEs and only one evolves in the MIR similarly to the variable ECLEs.
In \secref{sec:rate_analysis}, we calculate the rate at which \change{variable} ECLEs occur and compare them to observationally derived TDE rates.
We calculate the galaxy-normalized ECLE rate to be \galrate, the mass-normalized rate to be \massrate and the volumetric rate to be \volrate.
Our ECLE rates are consistent with emission from a subset of TDEs.
Finally, we list our conclusions in \secref{sec:conclusions}.

Throughout this paper, we assume a Hubble-Lema\^itre constant, $H_0$, of $\mathrm{73\ km\ s^{-1}\ Mpc^{-1}}$ and adopt a standard cosmology with $\Omega_\mathrm{m}=0.27$ and $\Omega_\Lambda=0.73$.

\section{Data}
\label{sec:data}
Here, we describe the instruments and surveys used to obtain the observations used in this work and the data-reduction and analysis techniques used to process the data.

\subsection{SDSS DR17}
\label{subsec:sdss_dr17}
We performed a search for ECLEs in the SDSS Legacy Survey DR17 \citep{DR172022}, which took observations between 2000 and 2008.
We used all spectra marked as `Galaxy' or `QSO' by the Survey.
Though we are most interested in variable ECLEs, we included QSOs in our sample to remain agnostic to the host galaxies.
\change{In addition, TDEs can create broad Balmer lines that may cause the SDSS pipeline to classify the galaxy as a QSO.
This is seen in the \citetalias{Wang2012} sample, where two of the variable ECLEs were classified as `QSO' by SDSS.}
We did not put a redshift limit on the galaxy sample, as the detection algorithm accounts for spectra where key emission lines lie outside the SDSS wavelength range.
For each spectrum, the only pixels used were those flagged as `good' (0) or `emission line' (40,000,000) by the SDSS pipeline.
If this cut removed 50 per cent or more of the pixels in a given spectrum, then the spectrum was not used.
This resulted in 13,664 spectra being rejected, which left a sample of 851,854 galaxies.

We also required each galaxy to have a stellar mass estimate, as these will be used later to calculate the rate at which ECLEs occur.
We preferentially used the galactic stellar masses derived by the MPA-JHU pipeline \citep{Kauffmann2003a,Brinchmann2004,Tremonti2004}.
Of the 851,854 galaxies, 82,298 did not have stellar masses in the MPA-JHU catalogue.
For these galaxies, we used the stellar mass values derived by the Portsmouth group pipeline \citep{Maraston2013}.
This pipeline fitted all galaxy spectra in the SDSS legacy Survey with two templates; a luminous red galaxy model and a star forming model.
The masses were then assigned using the parameters from the best fitting model.
The MPA-JHU and Portsmouth catalogues of galaxy properties have been shown to be consistent with each other \citep{Maraston2013,Thomas2013}.
3,269 galaxies did not have a stellar mass in either catalogue, so we removed them from our sample.
This resulted in a final sample of 848,585 galaxies.

\subsection{Optical spectroscopy}
\label{subsec:op_spec}
We obtained follow-up optical spectra of \change{seven} of the new ECLEs using the Dark Energy Spectroscopic Instrument (DESI) mounted on the Mayall 4 m telescope \citep{DESIcollab2016a,DESIcollab2016b,DESIcollab2023a,DESIcollab2023b}.
These spectra were observed as part of the Bright Galaxy Survey \citep{Hahn2023a} during main survey operations and were processed using the custom DESI spectroscopic pipeline \citep{Guy2023}.
The spectra were taken between March 2022 and April 2023.
It is important to note that SDSS and DESI have different sized fibres, with diameters of 3 and 1.5 arcsec, respectively \citep{Gunn2006,Abareshi2022}.
As a result, DESI spectra contain less light from the outer regions of the host galaxies despite being centred on the same location.
This may introduce relative changes in line fluxes and ratios depending on the line-emitting regions included or excluded by the fibres.

We also obtained optical spectra of \change{three} of the new ECLEs using the Gemini Multi-Object Spectrograph \citep[GMOS;][]{Hook2004} on the 8.1 m Gemini North Telescope (Gemini) on Maunakea, Hawai`i \change{as part of the Gemini programs GN-2023A-Q-322 and GS-2024A-Q-323 (PI: P. Clark)}.
These were taken on 2023 July 29 and 30, \change{and 2024 June 24 and 28} in the long-slit spectroscopy mode with a slit width of 1.0 arcsec, using the B480 and R831 gratings.
Data were reduced using the \verb|DRAGONS| (Data Reduction for Astronomy from Gemini Observatory North and South) reduction package \citep{Labrie2019}, using the standard recipe for GMOS long-slit reductions.
This includes bias correction, flatfielding, wavelength calibration, and flux calibration.
\change{\changetwo{As we did not have telluric standards for these observations, we use the python package}} \verb|TelFit| \change{\changetwo{\citep{Gullikson2014} to model and remove the telluric absorption features.}}

A summary of the observing conditions for these spectra are given in \tabref{tab:obs_conditions}.

\subsection{Optical photometry}
We obtained \textit{ugriz} photometry of two targets using the IO:O instrument \citep{Barnsley2016} mounted on the Liverpool Telescope \citep[LT;][]{Steele2004}.
For each target, single-epoch observations were obtained to look for significant variation since the SDSS photometry.
Basic instrumental pre-processing was performed on the images using the IO:O pipeline.

The LT~\textit{ugriz} photometry was measured on the pre-processed images following median stacking using \textsc{ysfitsutil} \citep{Bach2023}.
Photometric extraction was then conducted using \textsc{AutoPhOT} \citep{Brennan2022} using a Moffat point spread function \citep{Moffat1969} with calibration to the SDSS magnitude system using the SDSS DR16 photometric catalogue \citep{Ahumada2020} retrieved from VizieR \citep{Ochsenbein2000}.\footnote{\url{https://vizier.cds.unistra.fr/}}

\subsection{Mid-infrared photometry}
\label{subsec:ir_phot}
We retrieved MIR photometry from \textit{WISE} for all the ECLEs in our sample using the ALLWISE \citep{Wright2010} and NEOWISE Reactivation \citep{Mainzer2011,Mainzer2014} data releases.
\change{We searched for \textit{WISE} sources within a 3 arcsec radius of the locations of the ECLEs and then processed the data retrieved.}
In order to study the long-term variability of the \textit{WISE} data, we processed it using a custom \verb|python| script \citep{Clark2024} that removes observations that were marked as an upper limit; or were taken when the spacecraft was close to the South Atlantic Anomaly or the sky position of the Moon, or were flagged by the \textit{WISE} pipeline as having a low frame quality or suffering from potential `contamination or confusion.' \citet{Dou2016} showed that the \citetalias{Wang2012} variable ECLEs did not show MIR variability during each observation block.
Therefore, a weighted average was used to produce a single magnitude value per filter for each observation block.
This allowed us to more easily study the long-term evolution of the ECLEs.

\section{CL galaxy Discovery and Classification}
\label{sec:CL_galaxies}
In this section, we outline the algorithm used to detect CLs in galaxy spectra, compare the sample of galaxies with detected CLs to the overall SDSS sample, and discuss the detection efficiency of our algorithm (the detection code will be described at length by Clark et al., in prep.).
ECLEs, which are a small subset of the larger sample of galaxies with detected CLs, are discussed in the next section.

\subsection{CL galaxy search}
\label{subsec:CL_search}
\change{We measured the equivalent widths (EWs) of a set of emission lines for every galaxy in our sample.
The emission lines measured were the CLs observed in the \citetalias{Wang2012} ECLEs described in \secref{sec:intro}, as well as \fspectralline{O}{iii}{5007}, H$\upalpha$, H$\upbeta$, and \fspectralline{N}{ii}{6584}.
These were used to select CL galaxies and ECLEs and produce diagnostic diagrams.}

\change{To measure the EW of the emission lines in the galaxy spectra, we first isolated a 250 \AA\ section of the spectrum around each emission line.
We then fit a linear continuum across the location of the emission line.
For the narrow emission lines, this was fit between $\pm50$~\AA\ of the lines' wavelengths.
This range was modified for some of the lines if another prominent line appeared near the central wavelength.
For the typically broader lines, this was done over a larger range to ensure that the fit was done to the continuum.
The section of the spectrum was then normalized using the continuum fit.
The EW and signal-to-noise ratio (SNR) of the emission line was calculated from this normalized section of the spectrum.}

Our detection algorithm selected CL galaxies by detecting a sufficient number of CLs or at least one extremely strong CL.
\change{We first required that the peak of the emission feature was within $350~\mathrm{km~s}^{-1}$ of the zero velocity point of the line.
We also required the emission feature to be the strongest feature within $800~\mathrm{km~s}^{-1}$ of the zero velocity point.
These checks ensured that we were detecting the correct emission line and that we were not mistaking noise for an emission feature.}
A CL was detected in a galaxy if its EW is $<-1.5$ \AA\ for \fspectralline{Fe}{vii}{6088}, [\ion{Fe}{x}], [\ion{Fe}{xi}], or [\ion{Fe}{xiv}] (which we consider primary lines), or $<-0.5$ \AA\ for \fspectralline{Fe}{vii}{3759}, \fspectralline{Fe}{vii}{5160}, and \fspectralline{Fe}{vii}{5722} (which we consider secondary lines).
We assigned the detection of a primary line a score of two and a secondary line a score of one.
For a galaxy to be classified as a possible ECLE, its total score had to be five or greater.
If the galaxy was at a redshift where the [\ion{Fe}{xi}] line falls outside the wavelength range of the SDSS spectrograph, then this threshold was lowered to three to account for the line not being detectable.
We also flagged galaxies that exhibit at least one CL with an SNR of $>10$ or two of the four [\ion{Fe}{vii}] lines have SNRs $>5$.
These detection criteria are motivated by the fact that the \citetalias{Wang2012} sample all showed strong CLs from [\ion{Fe}{vii}], [\ion{Fe}{x}], [\ion{Fe}{xi}], and [\ion{Fe}{xiv}].
Using the detection algorithm on the sample of 848,585 galaxies and QSOs returned 969 CL galaxies.

\begin{figure}
    \centering
    \includegraphics[width=0.45\textwidth]{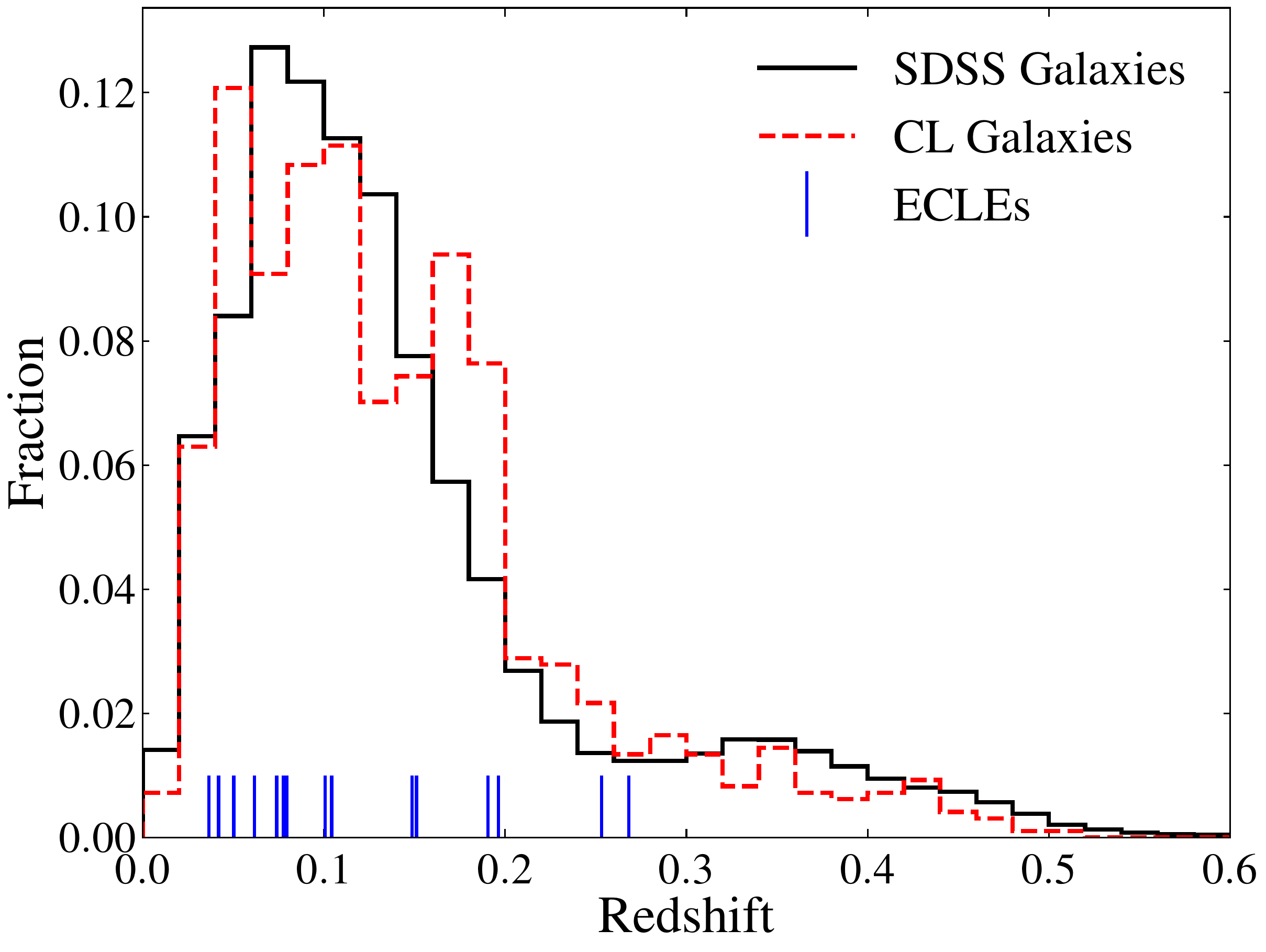}
    \caption{Redshift distributions of the SDSS (solid black curve) and CL (dashed red curve) galaxy samples.
    The redshifts of the \change{16} ECLEs in the galaxy sample are shown by the vertical blue lines.
    The distributions are broadly consistent but the CL galaxies are over-represented at $z\sim0.05$ and $z\sim0.18$ at 99 per cent confidence.}
    \label{fig:z_hist}
\end{figure}

\begin{figure}
    \centering
    \includegraphics[width=0.45\textwidth]{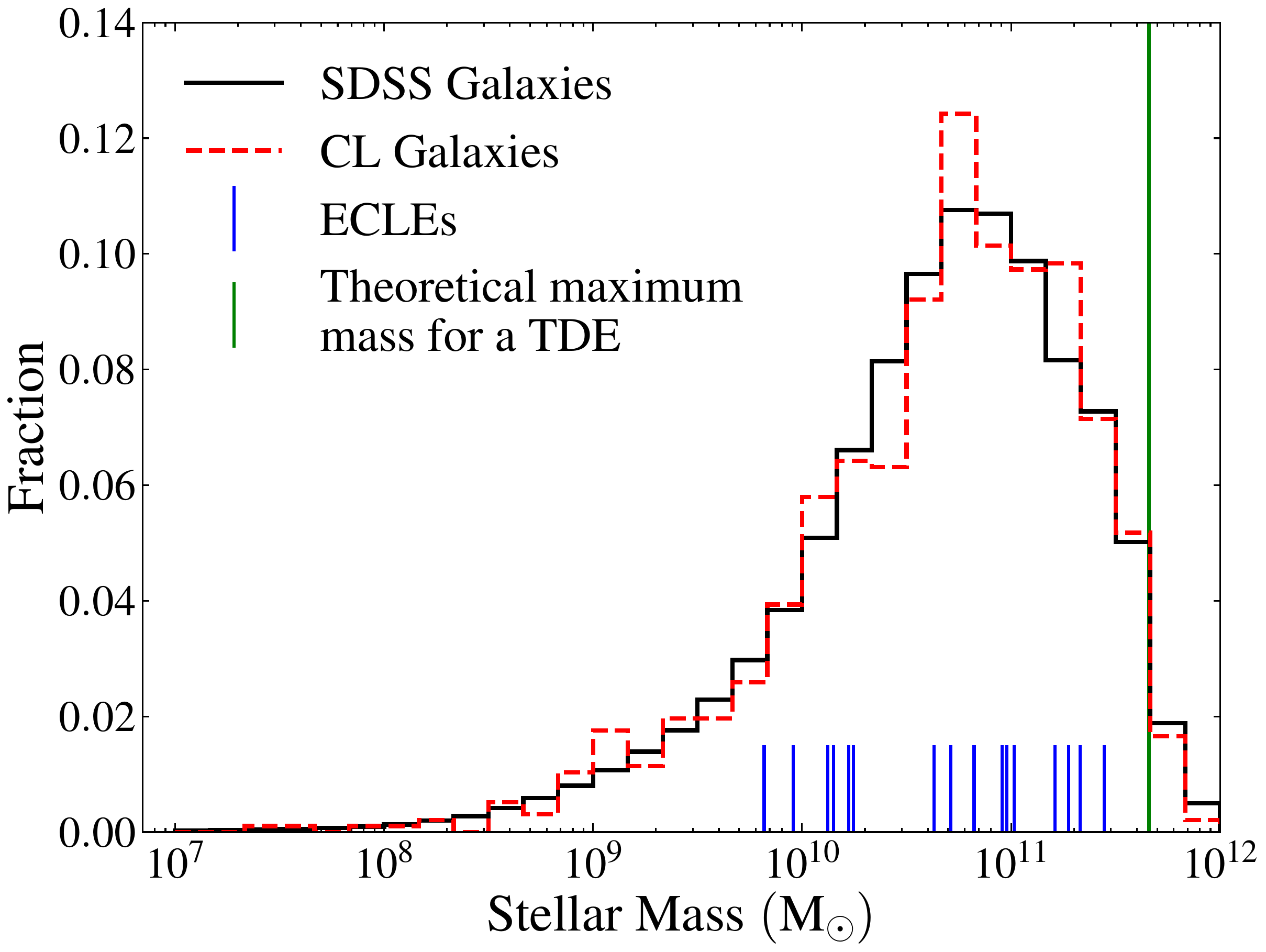}
    \caption{Stellar mass distributions of the SDSS (solid black curve) and CL (dashed red curve) galaxy samples.
    The masses of the 14 ECLEs in the galaxy sample are shown by the vertical blue lines.
    The vertical green line indicates the theoretical galaxy stellar mass limit above which a Sun-like star would fall directly into the galaxy's SMBH instead of being disrupted as a TDE \citep{Rees1988}.
    The relation between the stellar masses of the galaxy and its SMBH used in this calculation is from \citet{Reines2015}.}
    \label{fig:mass_hist}
\end{figure}

\begin{figure*}
    \centering
    \includegraphics[width=0.9\textwidth]{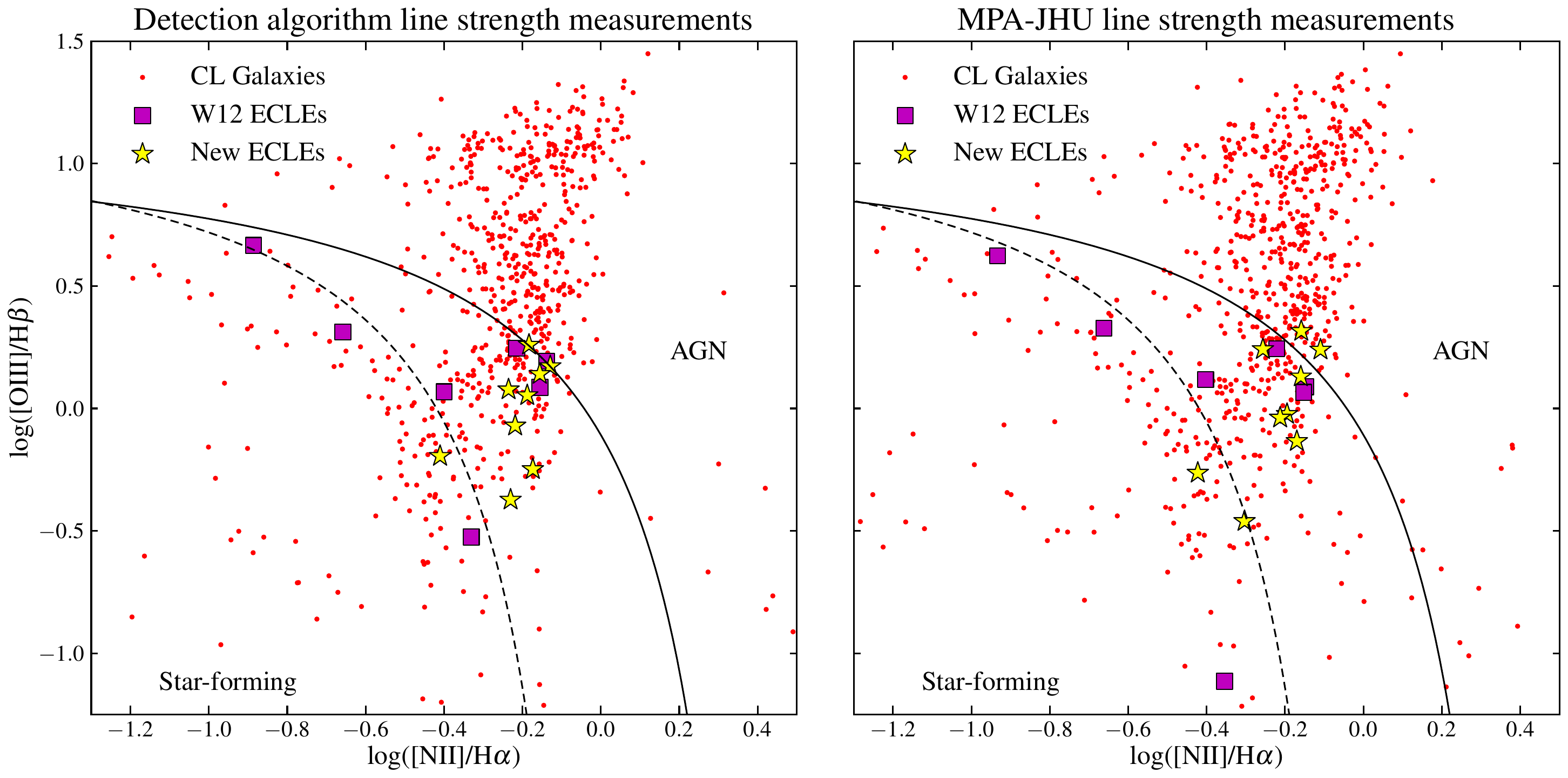}
    \caption{\change{A comparison of BPT diagrams created using the line strengths measured by our detection algorithm (left) and the MPA-JHU catalogue (right).
    The CL galaxies detected are shown as red circles, with the \citetalias{Wang2012} and new ECLEs shown as magenta squares and yellow stars, respectively.
    AGN are found above the solid curve \citep{Kewley2006} while star-forming galaxies are found below the dashed curve \citep{Kauffmann2003b}.
    The region between the star-forming and AGN regions is known as the composite region.
    The galaxies all lie in similar areas in both plots, so we are confident in using our line strengths going forward.}}
    \label{fig:bpt_comp}
\end{figure*}

Initially, we compared the CL galaxies to the overall SDSS sample.
\figref{fig:z_hist} shows the redshift distributions of the two samples.
Overall, both have similar shapes and a median redshift of 0.12.
However, a two-sided Kolmogorov-Smirnov test allows us to reject the null hypothesis that the two distributions are drawn from the same population with a \textit{p}-value $< 0.01$.
Despite having similar overall shapes, this difference is most notable at redshifts below 0.2 where the CL galaxies are over-represented at redshifts $\sim$0.05 and $\sim$0.18.
At \change{particular} redshifts in these bins, the wavelengths of [\ion{Fe}{X}] and [\ion{Fe}{XI}] coincide with the rest wavelengths of emission lines created by skylines.
\change{This did not affect all the galaxies in these redshift bins, so we did not remove galaxies from these bins from our sample.}
Galaxies with these skylines were removed during the visual inspection stage.

We also investigated the stellar masses of the galaxies selected by our algorithm.
\figref{fig:mass_hist} compares the galactic stellar mass distributions of the CL galaxies to the SDSS sample.
Again, the two distributions have similar shapes, with median masses $5.8\times10^{10}~\mathrm{M}_\odot$ and $5.4\times10^{10}~\mathrm{M}_\odot$ for the CL galaxies and SDSS sample, respectively.
\change{A two-sided Kolmogorov-Smirnov test does not allow us to reject the null hypothesis that the two distributions are drawn from the same population with a \textit{p}-value $> 0.01$.
Therefore, the CL galaxies have been selected across the full mass range of SDSS Legacy.}

\change{To test our method of measuring emission line strengths, we compared our detection algorithm to the MPA-JHU catalogue of SDSS-derived galaxy properties \citep{Kauffmann2003a,Brinchmann2004,Tremonti2004}\footnote{\url{https://wwwmpa.mpa-garching.mpg.de/SDSS/}} using \citet*[BPT]{Baldwin1981} diagrams.
The MPA-JHU pipeline modelled galaxy spectra using a library of template spectra, while taking into account the star formation histories of each galaxy.
\figref{fig:bpt_comp} shows the BPT diagrams used to test our line strength measurements.
By eye, the majority of the CL galaxies occupy a similar region of the plot for both methods.
Additionally, the \citetalias{Wang2012} ECLEs are in very similar positions in both.
In the plot made using our line strength measurements, 61 per cent of the CL galaxies lie in the AGN region of the plot, compared to 59 per cent for the MPA-JHU measurements.
Given the apparent similarities between the two plots, we are confident in using our line strength method to detect and measure CLs.}

\subsection{Detection efficiency}
\label{subsec:det_eff}
In order to determine the rate at which ECLEs occur, we needed to establish how many ECLEs we may have missed.
Our detection algorithm may have missed ECLEs for a variety of reasons, including masked areas covering the CLs or a low SNR for the whole spectrum.
We determined the efficiency of our detection algorithm by running it on fake ECLE spectra generated by planting CLs in a random sample of real SDSS Legacy spectra.

10,000 spectra were randomly selected from the galaxy and QSO sample described in \secref{subsec:sdss_dr17}.
For each spectrum, the CLs from a randomly selected ECLE from the \citetalias{Wang2012} sample were planted at the CLs' wavelengths.
To explore what effect the strength of the CLs had on the detection efficiency, a scaling factor was applied to all the CLs in each spectrum.
This scaling factor was randomly sampled from between 0 and a maximum value set depending on the presence of \fspectralline{O}{iii}{5007} in the base spectrum.
If this line was present, then the maximum scaling factor was set such that the strongest CL would have the same strength as [\ion{O}{iii}].
If it was not present, the maximum scaling factor was set to 1.
This was motivated by the fact that none of the CLs in the \citetalias{Wang2012} ECLEs was stronger than the \fspectralline{O}{iii}{5007} line in the spectra if it was present.

\begin{figure}
    \centering
    \includegraphics[width=0.45\textwidth]{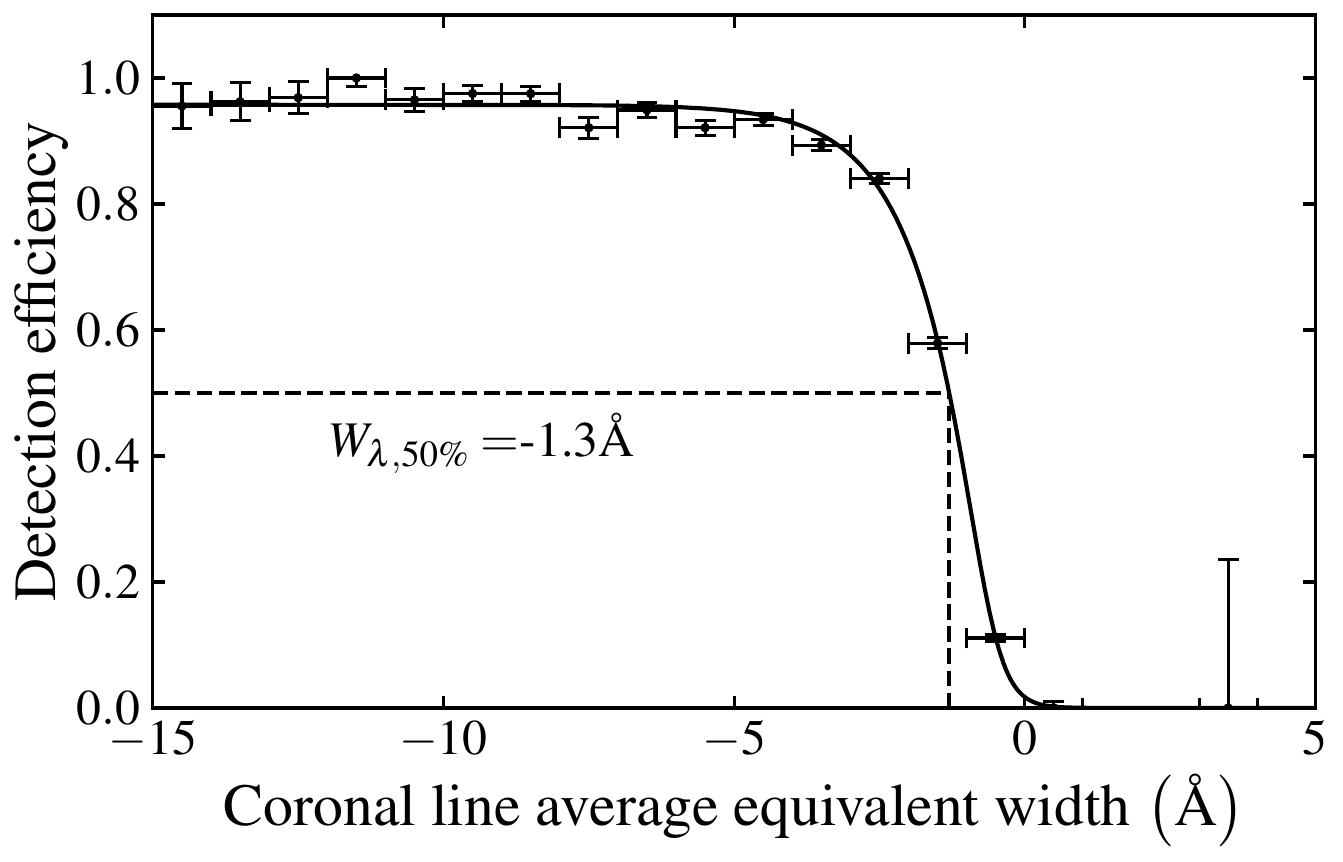}
    \caption{ECLE detection efficiency as a function of the average EW of the CLs.
    Points denote the fraction of fakes classified as ECLEs in 1 \AA\ bins.
    The curve is a generalized sigmoid fit to the data.
    Error bars indicate 1$\upsigma$ binomial uncertainties.}
    \label{fig:det_eff_curve}
\end{figure}

The detection algorithm was then run on the sample of fake ECLE spectra.
The results are shown in \figref{fig:det_eff_curve}, where we show the detection efficiency as a function of the average EW of the CLs.
Our maximum detection efficiency is 95 per cent and we reach 50 per cent detection efficiency at an average CL EW of -1.3 \AA.

\section{ECLE Sample}
\label{sec:ecle_sample}
In this section, we analyse the sample of ECLEs we have discovered in the SDSS Legacy Survey DR17.
We describe the selection criteria used to select the ECLEs out of the larger sample of CL galaxies and compare them to the \citetalias{Wang2012} sample.
We then describe each of the new ECLEs in turn and explore their properties using MIR data, follow-up spectra, and optical photometric observations.

\begin{table*}
    \centering
    \caption{Summary of the ECLEs detected in the SDSS Legacy Survey DR17.
The top section contains the new ECLEs from this work, while the bottom contains the ECLEs from \citetalias{Wang2012}.}
    \begin{tabular}{|c|c|c|c|c|c|}
        \hline
        SDSS Name & Short Name & RA (J2000) & Dec (J2000) & Redshift & Coronal Lines Present \\
        \hline
        \multicolumn{6}{c}{\bf{New ECLEs}} \\
        SDSS J080727.31+140537.0 & SDSS J0807 & 08:07:27.3157 & +14:05:37.0892 & 0.0738 & \shortfspecline{Fe}{vii}, \shortfspecline{Fe}{x} \\
        SDSS J120719.81+241155.8 & SDSS J1207 & 12:07:19.8102 & +24:11:55.8789 & 0.0503 & \shortfspecline{Fe}{vii}, \shortfspecline{Fe}{x}, \shortfspecline{Fe}{xi} \\
        \change{SDSS J123829.58+185237.5} &  \change{SDSS J1238} &  \change{12:38:29.5894} &  \change{+18:52:37.5554} &  \change{0.253} &  \change{\shortfspecline{Fe}{vii}} \\
        SDSS J124726.37+070525.0 & SDSS J1247 & 12:47:26.3719 & +07:05:25.0809 & 0.104 & \shortfspecline{Fe}{vii}, \shortfspecline{Fe}{x} \\
        SDSS J140204.75+293946.8 & SDSS J1402 & 14:02:04.7560 & +29:39:46.8759 & 0.196 & \shortfspecline{Fe}{vii} \\
        \change{SDSS J145849.72+191033.5} & \change{SDSS J1458} & \change{14:58:49.7267} & \change{+19:10:33.5109} & \change{0.268} & \change{\shortfspecline{Fe}{vii}, \shortfspecline{Fe}{x}} \\
        SDSS J145926.06+404538.5 & SDSS J1459 & 14:59:26.0676 & +40:45:38.5508 & 0.151 & \shortfspecline{Fe}{vii} \\
        SDSS J171504.28+564715.8 & SDSS J1715 & 17:15:04.2893 & +56:47:15.8404& 0.191 & \shortfspecline{Fe}{vii} \\
        SDSS J222055.73-075317.8 & SDSS J2220 & 22:20:55.7312 & -07:53:17.8464 & 0.149 & \shortfspecline{Fe}{vii} \\
        \hline
        \multicolumn{6}{c}{\bf{\citetalias{Wang2012} ECLEs}} \\
        SDSS J074820.66+471214.2 & SDSS J0748 & 07:48:20.6668 & +47:12:14.2648 & 0.0616 & \shortfspecline{Fe}{x}, \shortfspecline{Fe}{xi}, \shortfspecline{Fe}{xiv} \\
        SDSS J093801.63+135317.0 & SDSS J0938 & 09:38:01.6376 & +13:53:17.0423 & 0.101 & \shortfspecline{Fe}{vii}, \shortfspecline{Fe}{x}, \shortfspecline{Fe}{xi} \\
        SDSS J095209.56+214313.2 & SDSS J0952 & 09:52:09.5629 & +21:43:13.2979 & 0.0795 & \shortfspecline{Fe}{vii}, \shortfspecline{Fe}{x}, \shortfspecline{Fe}{xi}, \shortfspecline{Fe}{xiv} \\
        SDSS J105526.41+563713.1 & SDSS J1055 & 10:55:26.4177 & +56:37:13.1010 & 0.0740 & \shortfspecline{Fe}{vii}, \shortfspecline{Fe}{x}, \shortfspecline{Fe}{xi} \\
        SDSS J124134.25+442639.2 & SDSS J1241 & 12:41:34.2561 & +44:26:39.2636 & 0.0419 & \shortfspecline{Fe}{vii}, \shortfspecline{Fe}{x}, \shortfspecline{Fe}{xi}, \shortfspecline{Fe}{xiv} \\
        SDSS J134244.41+053056.1 & SDSS J1342 & 13:42:44.4150 & +05:30:56.1451 & 0.0365 & \shortfspecline{Fe}{x}, \shortfspecline{Fe}{xi}, \shortfspecline{Fe}{xiv} \\
        SDSS J135001.49+291609.6 & SDSS J1350 & 13:50:01.4946 & +29:16:09.6460 & 0.0777 & \shortfspecline{Fe}{x}, \shortfspecline{Fe}{xi}, \shortfspecline{Fe}{xiv} \\
        \hline
    \end{tabular}
    \label{tab:sample}
\end{table*}

\subsection{Cuts on CL galaxy sample}
\label{subsec:cuts}
To distinguish galaxies with strong CLs from those with weak lines, we followed the selection criteria suggested by \citetalias{Wang2012}.
For each spectrum, at least one CL had to be detected with an SNR of $>5$ and the strength of at least one CL had to be more than 20 per cent of the strength of the \fspectralline{O}{iii}{5007} line in that spectrum.
This was motivated by work showing that the strengths of CLs created by AGN are typically only a few per cent of their \fspectralline{O}{iii}{5007} line \citep{Nagao2000}.
These criteria reduced our sample to \change{332} galaxies.

\change{We note that we did not remove galaxies from our sample that could be classified as AGN using our BPT diagram (\figref{fig:bpt_comp}).
\citet{Clark2024} showed that four of the \citetalias{Wang2012} variable ECLEs migrated to the AGN side of the BPT diagram in follow-up observations, due to changes in the emission line strengths.
Therefore, to ensure we did not remove variable ECLEs that look like AGN on diagnostic diagrams, we retained all of these galaxies.}

Finally, we visually inspected the remaining candidates to remove false positives detected because of random noise or coincidence of skylines with the locations of CLs.
This reduced the sample to \change{16} ECLEs, including the seven ECLEs detected by \citetalias{Wang2012}.
The properties of the \change{16} ECLEs are shown in \tabref{tab:sample}.

\subsection{Sample comparisons}
\label{subsec:samp_comp}
There are some comparisons that can be made between the \citetalias{Wang2012} sample and our \change{nine} new ECLEs.
First, the majority of our new ECLEs are at higher redshifts than the \citetalias{Wang2012} sample (at a median redshift of \change{0.16}, compared to the \citetalias{Wang2012} median of 0.067).
This may be due to the SDSS Legacy spectra being reprocessed in DR8, after the \citetalias{Wang2012} search, allowing us to detect \change{strong CLs in noisier spectra}.
In addition, all \change{nine} of our new ECLEs were classified as `QSOs' by SDSS, whereas four of the \citetalias{Wang2012} sample were classified as `Galaxies'.

Another difference between the \citetalias{Wang2012} sample and our new ECLEs is the ionization states of the CLs.
The \citetalias{Wang2012} sample shows a range of CLs, with \shortfspecline{Fe}{x} appearing in every object and all but one object showing at least two lines of the three species \fspectralline{Fe}{vii}{6088}, \shortfspecline{Fe}{xi}, or \shortfspecline{Fe}{xiv}.
In comparison, the seven new ECLEs in our sample were detected for having sufficiently strong \shortfspecline{Fe}{vii} lines.
\shortfspecline{Fe}{x} was present in \change{four} ECLEs in the new sample and \shortfspecline{Fe}{xi} was only present in one (SDSS 1207).
This suggests that for the majority of our ECLEs, the processes creating the emission lines are at lower energies than for the \citetalias{Wang2012} sample.
This could be due to the TDEs producing the lines being more evolved than those in the \citetalias{Wang2012} sample.
As the star's matter is consumed by the SMBH, the luminosity of the flare decreases, so there would then be insufficient energy to create higher-ionization CLs.
This is seen in follow-up observations of the \citetalias{Wang2012} sample \citep{Yang2013,Clark2024}.
It is also possible that the lines are created by a non-variable, lower-energy progenitor, such as an AGN.
\change{\figref{fig:bpt_comp} shows that most of the new ECLEs fall within the composite region between the \citet{Kewley2001} and \citet{Kauffmann2003b} classification lines, which makes them liable to harbour AGN activity.}

\subsection{Notes on individual ECLEs}
\label{subsec:ind_objs}
\begin{figure*}
    \centering
    \includegraphics[width=\textwidth]{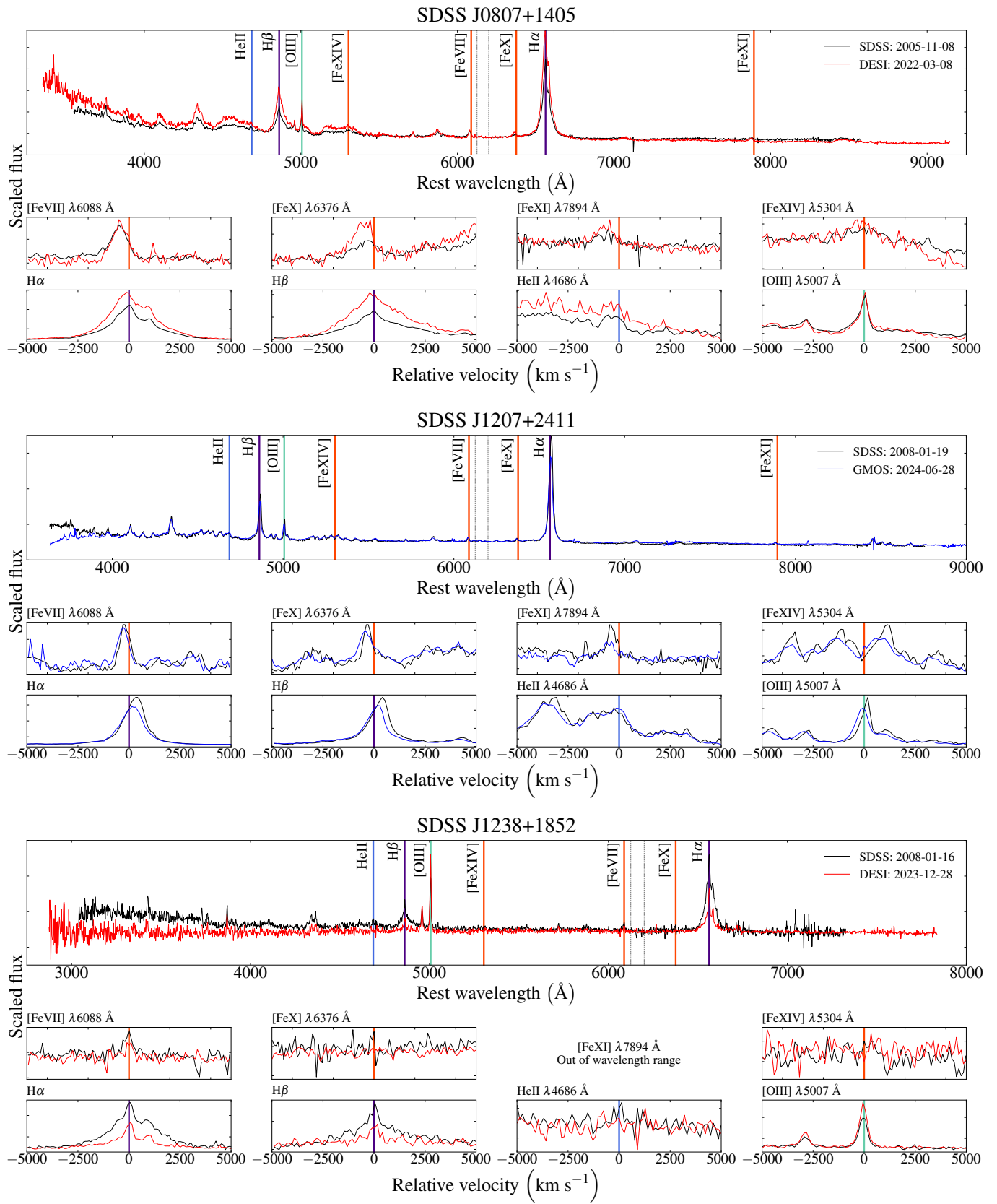}
    \caption{Comparisons of the new ECLE spectra from SDSS (black), DESI (red), and GMOS (blue).
    Emission lines of interest are marked by vertical lines.
    The dotted lines in the upper plots indicate the region used to rescale the spectra with respect to each other to allow for easy comparison.
    For the line-specific plots, this rescaling was done on continuum sections of the spectra near the emission line.}
    \label{fig:spec_mosaic}
\end{figure*}

\begin{figure*}
    \centering
    \includegraphics[width=\textwidth]{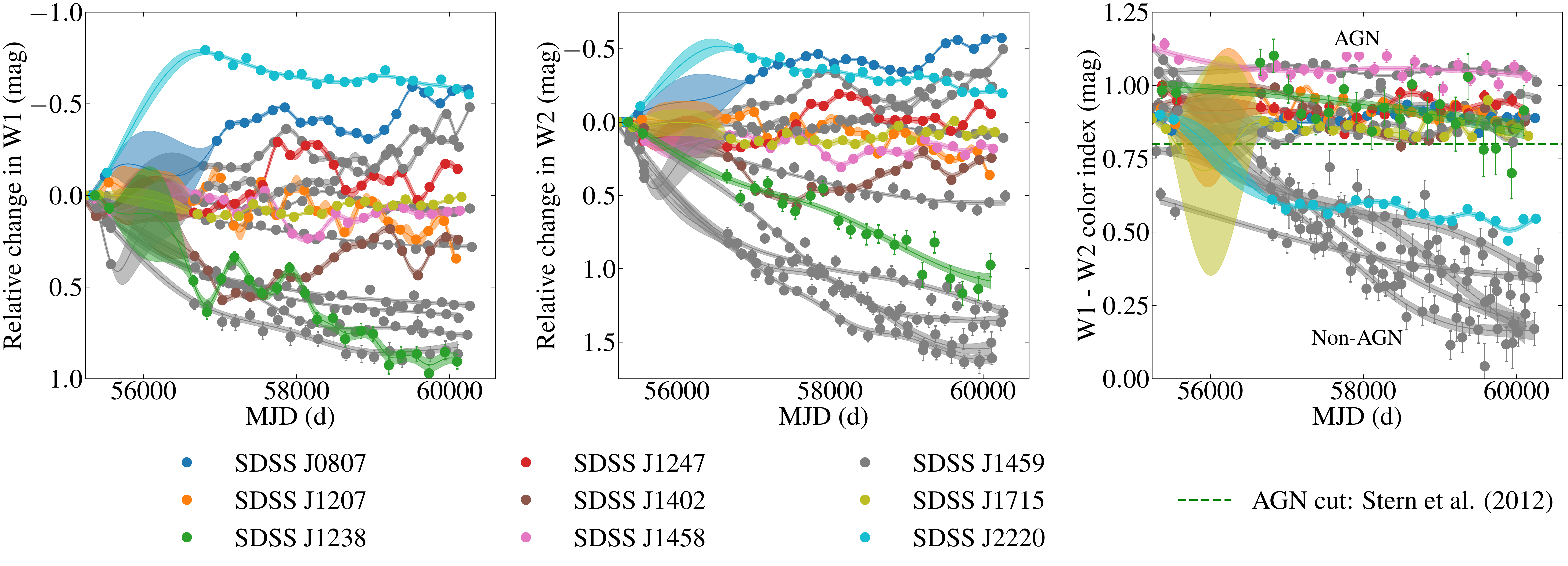}
    \caption{Comparison of the MIR evolution of new ECLEs (in colour) and the \citetalias{Wang2012} sample (in grey).
    The left and centre panels show the evolution in the W1 and W2 bands.
    The right panel shows the colour evolution.
    The dashed horizontal line is the AGN/non-AGN dividing line from \citet{Stern2012}.
    \change{SDSS J1238 is the only new ECLE that evolves in the individual bands similarly to the variable ECLEs from the \citetalias{Wang2012} sample.
    However, its colour index remains above the AGN dividing line.}}
    \label{fig:mir_mosaic}
\end{figure*}

\begin{figure*}
    \centering
    \includegraphics[width=0.9\textwidth]{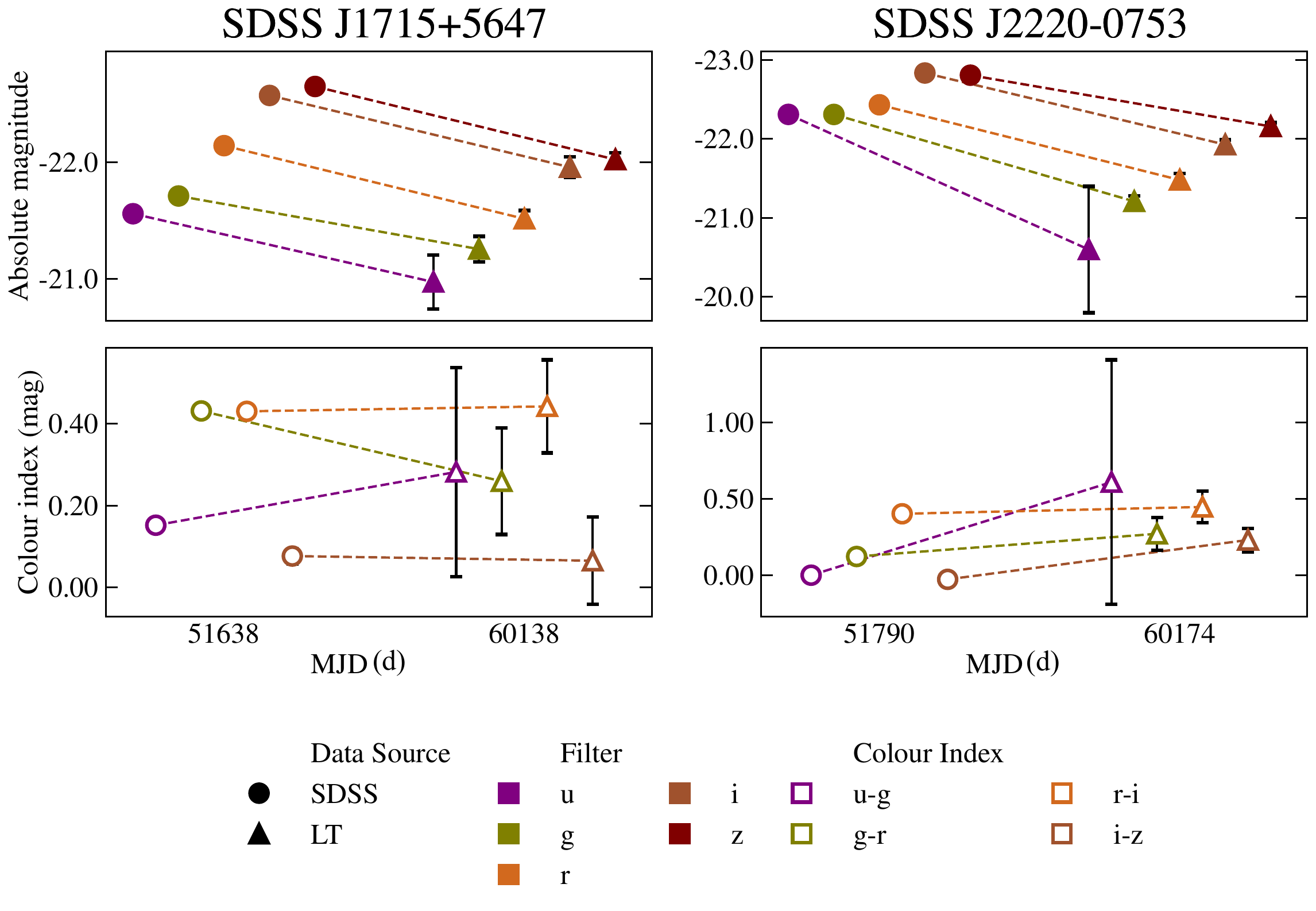}
    \caption{Optical photometric comparisons of SDSS J1715 and SDSS J2220 between the SDSS photometric observations and recent observations made with the LT.
    The left panel of each plot shows the SDSS \textit{ugriz} observations and the right panels shows the observations made with the LT.
    In each panel, the points have been shifted horizontally for clarity.
    The LT observations of SDSS J1715 show that it has dimmed by $\sim0.5$ mag since the SDSS observation, but the colour has not changed.
    However, SDSS J2220 has dimmed by $\sim1$ mag and has become bluer since the SDSS observation.}
    \label{fig:phot_evo}
\end{figure*}

In this section, we describe each of the \change{nine} new ECLEs in our sample in turn and explore their properties using follow-up spectra (\figref{fig:spec_mosaic}), MIR data (\figref{fig:mir_mosaic}), and optical photometric observations (\figref{fig:phot_evo}).
Only SDSS J1207 and SDSS J1715 do not currently have DESI spectra.
The details of the DESI spectra for the other ECLEs are listed in \tabref{tab:desi_specs}.
When discussing the MIR colour evolution, we use the $W1-W2\ge0.8$ AGN dividing line to classify AGN MIR activity \citep{Stern2012,Assef2013}.
We searched for previous transient activity in the \change{nine} new ECLEs by cross-matching them with the Transient Name Server.\footnote{https://www.wis-tns.org/}
None of the \change{nine} ECLEs were reported as having transient activity in the last several years.
This lack of reports supports our assumption that if the CLs were created by a variable progenitor, then it was a single-epoch event and not a recurring process.

\begin{Contfigure*}
    \centering
    \includegraphics[width=\textwidth]{images/SDSS_vs_DESI_Comparison_Mosaic_2.pdf}
    \caption{Comparisons of the new ECLE spectra from SDSS (black), DESI (red), and GMOS (blue).
    Emission lines of interest are marked by vertical lines.
    The dotted lines in the upper plots indicate the region used to rescale the spectra with respect to each other to allow for easy comparison.
    For the line-specific plots, this rescaling was done on continuum sections of the spectra near the emission line.}
\end{Contfigure*}

\begin{Contfigure*}
    \centering
    \includegraphics[width=\textwidth]{images/SDSS_vs_DESI_Comparison_Mosaic_3.pdf}
    \caption{Comparisons of the new ECLE spectra from SDSS (black), DESI (red), and GMOS (blue and magenta).
    Emission lines of interest are marked by vertical lines.
    The dotted lines in the upper plots indicate the region used to rescale the spectra with respect to each other to allow for easy comparison.
    For the line-specific plots, this rescaling was done on continuum sections of the spectra near the emission line.}
\end{Contfigure*}

\subsubsection{SDSS J0807+1405}
SDSS J0807 was selected for having strong \shortfspecline{Fe}{vii} lines, but visual inspection revealed the presence of weak \shortfspecline{Fe}{x} as well.
The latter was not detected by the search algorithm as it did not have a large enough EW.
The spectrum has strong, broad Balmer features and strong \fspectralline{O}{iii}{5007}.
Comparing the SDSS spectrum to the DESI spectrum, a major difference is that the DESI spectrum \change{is possibly} bluer and the Balmer lines and \shortfspecline{Fe}{x} line strengths have increased.
The MIR evolution of SDSS J0807 does not show the characteristic decline of variable ECLEs.
Instead, it appears to show a gradual brightening in the MIR.
The colour index has remained constant over the observation time and above the AGN colour cut.
Altogether, this suggests that the source is somewhat variable and has been brightening over a long timescale.
\change{As the CLs have not faded and the MIR evolution is unlike that of the \citetalias{Wang2012} variable ECLEs, we conclude that SDSS J0807 is not a variable ECLE.
Given that the spectrum has possibly become bluer, the Balmer lines have strengthened, the MIR emission is brightening, and the galaxy's W1$-$W2 colour lies above the AGN dividing line, it is likely that this is an AGN which has undergone a period of increased accretion.}

\subsubsection{SDSS J1207+2411}
SDSS J1207 was selected for having strong \shortfspecline{Fe}{vii} lines, but when visually inspected was found to have weak \shortfspecline{Fe}{x} and \shortfspecline{Fe}{xi} lines as well, which were not detected due to being weaker than our detection threshold.
All the CLs have a similar velocity shift and so are likely to have been produced in the same region.
\change{Both the \fspectralline{Fe}{vii}{6088} and \shortfspecline{Fe}{x} lines have remained at the same strength in the GMOS spectrum.
The \shortfspecline{Fe}{xi} line has appeared to weaken, but given this was very weak in the SDSS spectrum, it is difficult to determine if this line was real to begin with.
In addition, the MIR individual bands and colour evolutions have remained constant throughout the \textit{WISE} observations.
Therefore, we are confident that SDSS J1207 is not a variable ECLE.}

\change{\subsubsection{SDSS J1238+1852}
SDSS J1238 exhibits \shortfspecline{Fe}{vii} lines but no other CLs.
The DESI spectrum shows that the CLs have all weakened but are still present.
In addition, the Balmer lines have weakened and narrowed significantly, and the spectrum overall is possibly redder.
This indicates that the phenomenon powering the emission lines and continuum luminosity has dimmed in the 15 yr between the spectra.
However, the \fspectralline{O}{iii}{5007} line has slightly increased in strength, which does not appear to agree with this assessment.
The individual MIR bands show a decrease in magnitude over the observation period of \textit{WISE} in a similar manner to that of the variable ECLEs from \citetalias{Wang2012}.
However, the W1 band evolution is very erratic, in contrast to the smooth evolutions of the variable ECLEs.
Furthermore, the colour evolution of SDSS J1238 has remained above the AGN dividing line, and has also been erratic over the observation period.
It is possibly exhibiting a slight downward trend, but we are unable to confirm this at this time.
Given that the CLs have only weakened and not disappeared from the spectrum and the overall consistency of the MIR colour evolution, we do not consider SDSS J1238 to be a variable ECLE like the previous sample.
Given the variability in the Balmer lines and individual MIR bands, it is possible that this is a changing-state AGN with strong CLs that has been transitioning over the past 15 yr.
Further MIR follow-up is needed to confirm whether the colour index will eventually fall below the AGN dividing line.
}

\subsubsection{SDSS J1247+0705}
SDSS J1247 was selected for having strong \fspectralline{Fe}{vii}{6088}, but visual inspection revealed it also has weak \shortfspecline{Fe}{x}.
The DESI spectrum shows that all the CLs are still present and \fspectralline{O}{iii}{5007} has remained at the same strength  and the MIR colour evolution stays above the AGN dividing line.
Therefore, \change{SDSS J1247 is not a variable ECLE.}

\subsubsection{SDSS J1402+2939}
SDSS J1402 was selected for having strong \shortfspecline{Fe}{vii} lines but shows no other CLs.
The DESI spectrum shows that all the lines of interest have stayed at roughly the same strength.
Though there is some variation in the MIR brightness and colour evolution, the latter remains above the AGN dividing line.
\change{Therefore, SDSS 1402 is not a variable ECLE}.
One thing to note is that the DESI spectrum for this object appears to show a strong, narrow \fspectralline{Fe}{xi}{7894} emission line, which is appreciably narrower than the \fspectralline{Fe}{vii}{6088} line in the spectrum, i.e.
$\sim200~\mathrm{km~s^{-1}}$ compared to $\sim1200~\mathrm{km~s^{-1}}$.
While it is possible for different regions of circumnuclear material to produce CLs at different times, the declining luminosity of a TDE should not be able to produce lines at higher ionizations than those observed at earlier phases.
Hence, we consider this line to be an artefact.

\change{\subsubsection{SDSS J1458+1910}
SDSS J1458 was selected for having strong \shortfspecline{Fe}{vii} lines.
There also appeared to be \fspectralline{Fe}{x}{6376}, but on closer inspection this was found to be the result of a skyline being coincident with the CL location.
In the DESI spectrum, the \shortfspecline{Fe}{vii} lines have remained at the same strength.
In addition, the Balmer lines and \fspectralline{O}{iii}{5007} line profiles have remained the same.
The MIR brightness and colour evolution has also remained constant, with the W1$-$W2 colour lying above the AGN dividing line.
Therefore, SDSS J1458 is not a variable ECLE.}

\subsubsection{SDSS J1459+4045}
SDSS J1459 was selected for having moderately strong \shortfspecline{Fe}{vii} lines but shows no other CLs.
The CLs all remain at a similar strength in the DESI spectrum (\figref{fig:spec_mosaic}).
The MIR data show a decrease in brightness at a MJD of $\sim55,000$ d before brightening to its original magnitude and continuing to increase since.
During this time period, the colour evolution fluctuates but stays above the AGN dividing line (\figref{fig:mir_mosaic}).
\change{Therefore, SDSS J1459 is not a variable ECLE.}
One thing to note alongside the increase in both of the MIR bands is that the DESI spectrum shows that the object has become bluer since the SDSS spectrum.
Therefore, although the CLs are not variable, it appears SDSS J1459 is somewhat variable and has become brighter.

\subsubsection{SDSS J1715+5647}
SDSS J1715 was selected for having a strong \fspectralline{Fe}{vii}{6088} line and moderately strong secondary \shortfspecline{Fe}{vii} lines.
The SDSS spectrum also shows broad Balmer features and strong \fspectralline{O}{iii}{5007} emission, which suggests it is an AGN.
The GMOS spectrum is very similar to the SDSS spectrum, with all the emission lines of interest being very similar in strength and shape.
\change{The overall shape of the spectrum has changed slightly, with it possibly becoming more blue and less red.
There is no significant variation in the MIR observations and the colour index has remained above the AGN colour cut.}
Optical photometric observations taken with the LT \change{on 2023 July 13} show a decrease in magnitude of $\sim0.5$ mags in each SDSS band, indicating that the object has dimmed since the SDSS observation.
\change{The colour indices are broadly similar between the two observations, which indicates that the colour change in the spectra is not real.
As none of the CLs have faded, we consider it unlikely that SDSS J1715 is a variable ECLE.}

\subsubsection{SDSS J2220-0753}
SDSS J2220 was selected for having moderately strong \shortfspecline{Fe}{vii} lines but has no other CLs.
The SDSS spectrum exhibits broad Balmer features, which are common in AGN.
The DESI spectrum for this object shows that \fspectralline{Fe}{vii}{6088} has faded slightly and \fspectralline{O}{iii}{5007} has increased in strength, a behaviour observed in some \citetalias{Wang2012} variable ECLEs \citep{Yang2013,Clark2024}.
The Balmer features have stayed at the same strength but have narrowed slightly, indicating that the velocity of the broad line region has decreased.
\change{Both GMOS spectra show that the \fspectralline{Fe}{vii}{6088} line has continued to weaken.
It also appears that the Balmer lines have weakened slightly.
The first GMOS spectrum (blue) shows the \fspectralline{O}{iii}{5007} line at a similar strength as in the DESI spectrum.
However, in the second GMOS spectrum (magenta), it has faded and is weaker than in the original SDSS spectrum.
To investigate this further, we use the} \verb|python| package \verb|specutils| \change{to model the continua of the spectra and fit the \fspectralline{O}{iii}{5007} and \fspectralline{Fe}{vii}{6088} lines with single Gaussians.
We then measure the EWs and fluxes of these lines to confirm this variability.
These parameters are presented in \tabref{tab:j2220_lines}.
These show that both lines vary by roughly half an order of magnitude over the observation period.
}

\begin{table*}
    \centering
    \caption{\change{EWs and fluxes of the \fspectralline{O}{iii}{5007} and \fspectralline{Fe}{vii}{6088} emission lines in the spectra of SDSS J2220.
    GMOS 1 and GMOS 2 refer to the spectra taken on 2023-07-30 and 2024-06-24, respectively.
    Both the fluxes and EWs of both line vary significantly between the spectra, in a manner similar to that of the variable ECLEs.}}
    \begin{tabular}{|c|c|c|c|c|}
        \hline
        Spectrum & \shortfspecline{O}{iii} EW & \shortfspecline{O}{iii} Flux & \shortfspecline{Fe}{vii} EW & \shortfspecline{Fe}{vii} Flux \\
            & (\AA) & ($10^{-16}\ \mathrm{erg\ cm^{-2}\ s^{-1}}\ \textrm{\AA}^{-1}$) & (\AA) & ($10^{-16}\ \mathrm{erg\ cm^{-2}\ s^{-1}}\ \textrm{\AA}^{-1}$)\\
        \hline
        SDSS & $-5.8\pm0.5$ & $23.1\pm2.1$ & $-2.0\pm0.4$ & $5.2\pm1.0$ \\
        DESI & $-8.9\pm0.6$ & $37.6\pm1.8$ & $-1.8\pm0.4$ & $5.4\pm1.1$ \\
        GMOS 1 & $-11.5\pm0.6$ & $26.0\pm1.4$ & $-1.7\pm0.3$ & $2.8\pm0.5$ \\
        GMOS 2 & $-10.3\pm0.5$ & $19.7\pm1.0$ & $-1.0\pm0.3$ & $1.9\pm0.5$ \\
        \hline
    \end{tabular}
    \label{tab:j2220_lines}
\end{table*}

\begin{figure*}
    \includegraphics[width=0.235\textwidth]{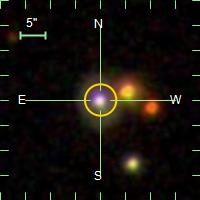}\hfil
    \includegraphics[width=0.27\textwidth]{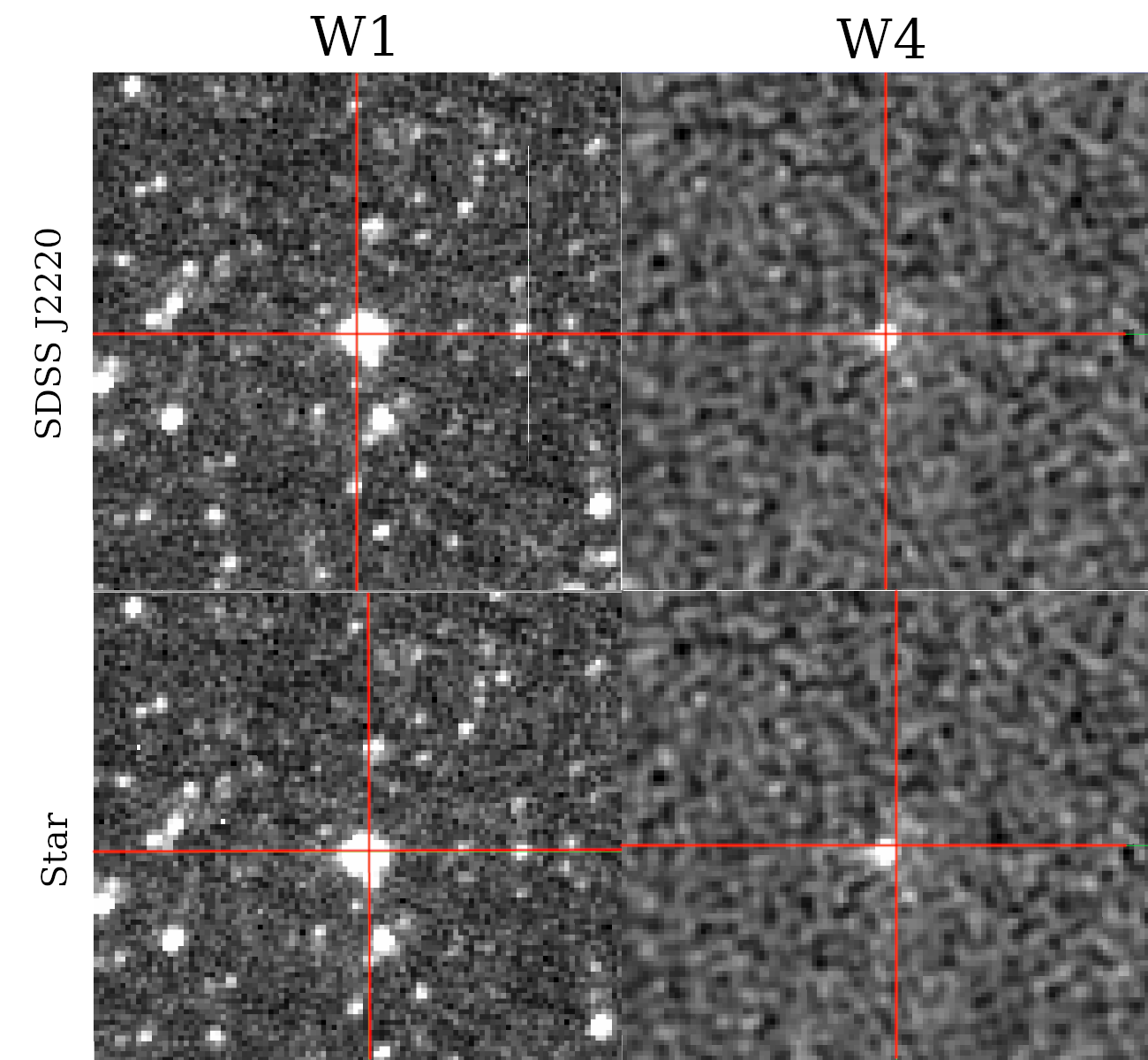} \\ [\smallskipamount]
    \includegraphics[width=\textwidth]{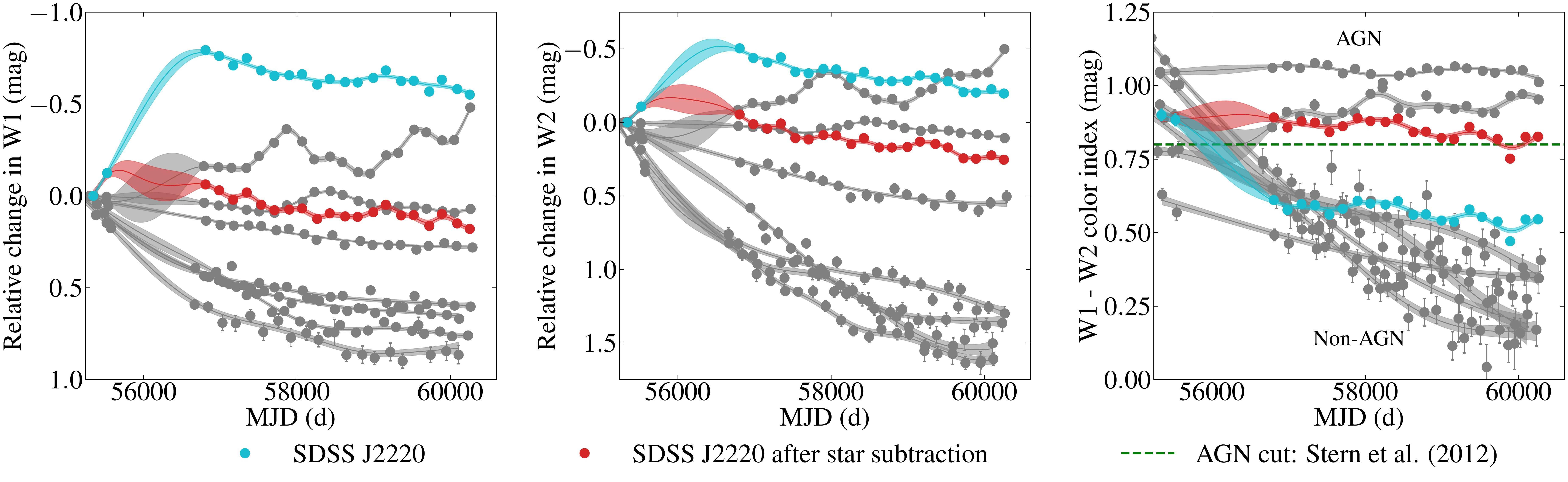}
    \caption{\change{\textit{Top left}: SDSS image of SDSS J2220 with the 3 arcsec radius circle used for retrieving \textit{WISE} data centred on the galaxy.
    The PSF of the \textit{WISE} satellite is $\sim6$ arcsec, which results in the MIR emission of SDSS J2220 and the star to the north-west appearing as one source.
    \textit{Top right}: Comparison of the W1 and W4 ALLWISE observations near SDSS J2220.
    In each panel, the red crosshairs show the RA and Dec of SDSS J2220 (\textit{top}) and the nearby star (\textit{bottom}).
    In the W1 band, the emission from the galaxy and the star are combined into one source.
    In the W4 band, the emission is centred on the coordinates of SDSS J2220, whereas the star shows no W4 emission.
    The W4 emission was used by ALLWISE to discern the two sources.
    As NEOWISE did not observe in W4, the emission from the two objects was not differentiated in the later releases.
    \textit{Bottom}: \changetwo{Comparison of the observed MIR evolution of SDSS J2220 (cyan, filled circles) to the MIR evolution with the contribution from the nearby star subtracted (cyan, empty circles), assuming the emission from the star is constant.
    We also include the full \textit{WISE} data of the \citetalias{Wang2012} sample (grey).
    Both the W1 and W2 bands of the star subtracted MIR emission decrease to below the initial magnitude.
    The W1$-$W2 colour also decreases slightly, but remains above the AGN dividing line.
    Compared to the \citetalias{Wang2012} sample, SDSS J2220 appears more similar to the non-variable ECLEs.}}}
    \label{fig:j2220_mir_summary}
\end{figure*}

\change{The MIR emission of SDSS J2220 is also unclear.}
Both MIR bands show a significant increase in magnitude followed by a slow decline over a period of $\sim3000$ d, which indicates variability but not in the same manner as expected for \change{variable ECLEs.}
In contrast, the colour evolution shows the transition from an AGN-like colour to a non-AGN-like colour, \change{similar to the variable ECLEs from the \citetalias{Wang2012} sample}.
\change{\figref{fig:j2220_mir_summary} shows the SDSS image of SDSS J2220, which has a number of other objects nearby, the closest being a bright star to the north-west.
We overlay the 3 arcsec radius used to select \textit{WISE} sources, which is sufficiently small to only include SDSS J2220.
In addition, the pipeline we use to process the \textit{WISE} data removed observations that were flagged as potentially contaminated.
However, the point spread function (PSF) of the \textit{WISE} satellite is $\sim6$ arcsec, which causes the galaxy and nearby star to merge into one source in the \textit{WISE} images.}

\begin{figure*}
    \centering
    \includegraphics[width=\textwidth]{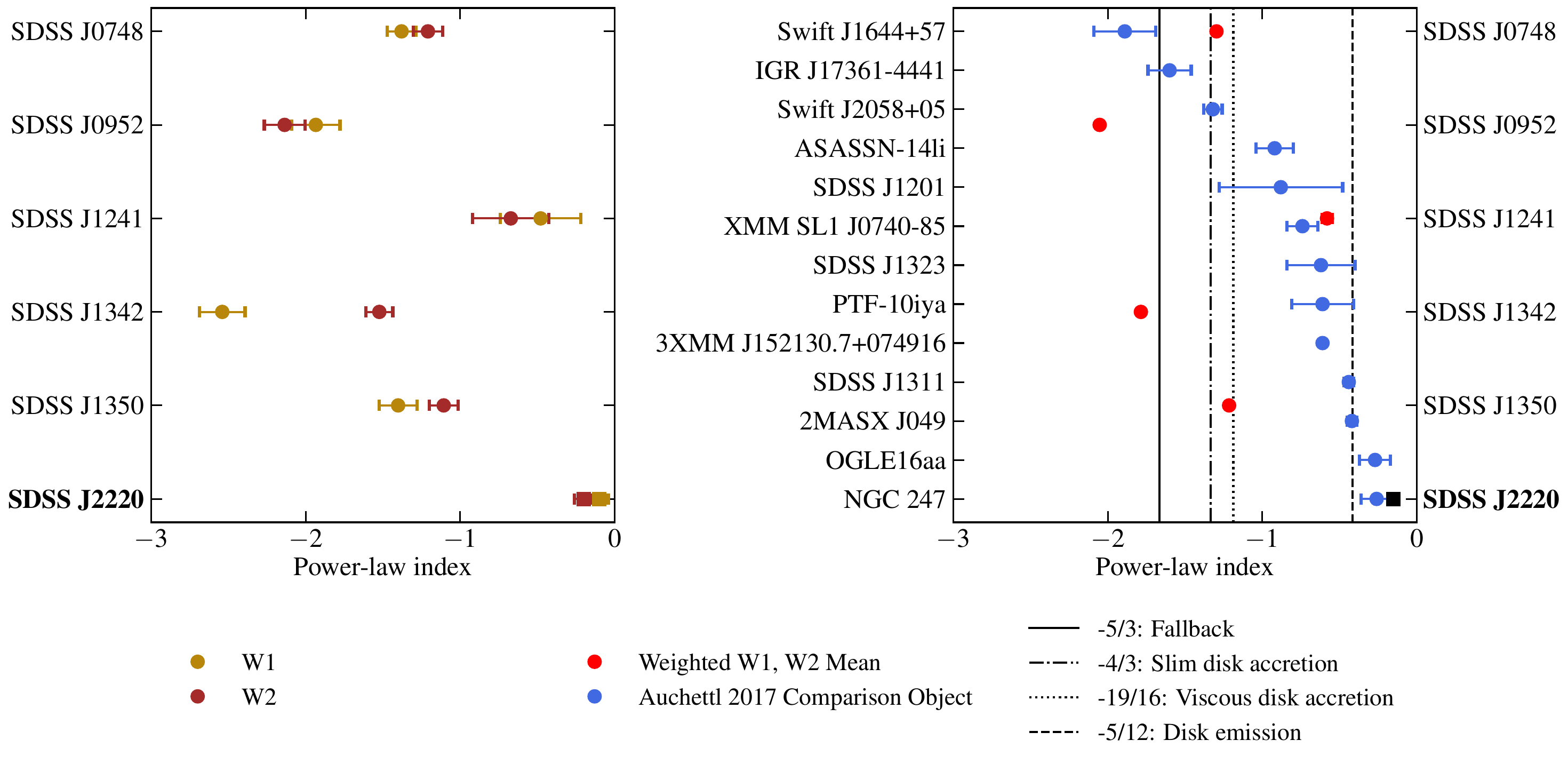}
    \caption{Comparison of the power law indices fit to the declining sections of the ECLE MIR data of SDSS J2220 and power law indices of X-ray TDE light curves from \citet{Auchettl2017}.
    \textit{Left}: Power law indices for the W1 and W2 band fits for the five variable ECLEs from the \citetalias{Wang2012} sample and the possible variable ECLE from this work, SDSS J2220.
    \textit{Right}: Weighted mean power law indices for the variable ECLEs and power law indices for the X-ray TDEs from \citet{Auchettl2017}.
    The weighted mean of the W1 and W2 indices for SDSS J2220 is consistent with both the \citetalias{Wang2012} sample and the X-ray TDEs.
    SDSS J2220 is represented by square markers in both plots.}
    \label{fig:index_comp}
\end{figure*}

\change{The reason for the large increase in brightness between the ALLWISE (the first two points of the MIR evolution before MJD 56,000) and NEOWISE (the remaining points) observations is due to a difference in the process used to distinguish sources between the two catalogues.
For the ALLWISE observations, the W4 band was also observed, which is produced by warm dust and so typically tracks star formation.
SDSS J2220 was bright in W4, whereas the nearby star showed no W4 emission (\figref{fig:j2220_mir_summary}).
With the aid of the W4 observations, the ALLWISE catalogue was able to separate the W1 and W2 emission of the two sources.
Therefore, the ALLWISE observations of SDSS J2220 are correct.
However, W4 was not observed in NEOWISE, so the sources could not be separated and the reported W1 and W2 magnitudes are a combination of SDSS J2220 and the nearby star.
The colour index of the star from the ALLWISE observations is $\sim0$, which when averaged with the SDSS J2220 colour index would give a value of $\sim0.5$, similar to what we see in the NEOWISE observations.
Given this contamination, we are unable to draw any firm conclusions about the nature of SDSS J2220 from the MIR increase in the W1 and W2 bands or the change in the colour index.
\changetwo{By assuming that the MIR emission from the nearby star is constant, we can subtract its contribution from the NEOWISE observations of SDSS J2220 to approximate the true MIR evolution of SDSS J2220 (\figref{fig:j2220_mir_summary}).
This removes the large jump in magnitude in both the W1 and W2 bands, as well as the decrease in colour.
All three panels show a much flatter evolution, though they all exhibit a slight decline.
The W1$-$W2 colour now remains above the AGN dividing line, more similar to the non-variable ECLEs.
}}

To explore the slight MIR decline, we fit the declining sections of the W1- and W2-band light curves with power laws, in the same manner as \citet{Clark2024}.
In \figref{fig:index_comp}, we compare the fitted power-law indices to the variable ECLEs from \citetalias{Wang2012} and a sample of TDEs compiled by \citet{Auchettl2017}.
We fit the W1-band with a free power-law index and the W2-band with both a free index and the fixed index from the W1 fit.
The W1-band fit gives a power-law index of $-0.10\pm0.06$, which is consistent with the \citetalias{Wang2012} sample.
The W2-band fit gives a power-law index of $-0.20\pm0.06$ when the index is free.
This is shallower than for any of the \citetalias{Wang2012} sample.
The weighted mean of these two indices is shallower compared to the \citetalias{Wang2012} sample and only falls within the errors of one index from the sample of X-ray TDEs from \citet{Auchettl2017}.

\change{The optical photometric evolution shows that the object has dimmed in each filter by roughly 1 mag, which is a significant change.
The colour evolution indicates that the object has largely remained the same colour, with some possible change in the u-g colour.
This is also seen in the spectral evolution, where the blue end has changed shape.
However, the variation in the spectra are difficult to assess, as the GMOS spectrum is more similar to the SDSS spectrum than the DESI one, despite being observed less than a year after the DESI spectrum.
It is possible that these variations are due to differences in the observing methods.
The LT observations were taken on 2023 August 18, $\sim20$ d after the first GMOS spectrum.}

\change{Overall, it appears that emission lines in SDSS J2220 are variable on a long timescale.
The CLs have dimmed but not disappeared over the $\sim20$ yrs between spectra, which is inconsistent with the \citetalias{Wang2012} variable ECLEs.
The contamination of the MIR emission by a nearby star makes it hard to draw strong conclusions about the MIR behaviour of this galaxy.
However, the MIR power law fits and increase in \shortfspecline{O}{iii} strength are consistent with the \citetalias{Wang2012} variable ECLE sample.
At this stage, we cannot tell conclusively whether or not SDSS J2220 is a variable ECLE.
As a result, we exclude it from the sample used for the rate calculation, but treat it as a +1 systematic uncertainty on the number of variable ECLEs in the sample.
We encourage further follow-up observations of this object to fully determine its nature.}

\change{
\subsection{ECLE selection criteria}
As shown by \figref{fig:bpt_comp}, though our method of measuring line strengths is similar to full spectral and Gaussian fitting, \changetwo{we may not be sensitive to small differences in strengths due to \shortspecline{Fe}{ii} complexes or the blending of adjacent features.}
This could affect whether the CLs appear to be stronger than 20 per cent of the strength of [O III], as required by our detection algorithm.
To investigate what effect this has, and the efficacy of the cut to remove AGN with CLs, we also inspect 77 galaxies in our CL sample that sport a CL-to-\shortfspecline{O}{iii} ratio between 10 and 20 per cent.
We visually inspected the SDSS spectra and MIR evolutions of these galaxies, selecting those with real CLs and non-AGN MIR evolutions.
Roughly 60 of the rejected galaxies were clear AGN with constant AGN-like MIR evolutions.
Therefore, we are confident that the ratio cut is effective at removing AGN from the CL galaxy sample.}

\change{Of the 77 galaxies, 14 had non-AGN MIR evolution, ten of which had DESI spectra, which we inspected for variable CLs.
Only two had possible CL variation, which when combined with the four without DESI spectra, leaves six galaxies with strong CLs and non-AGN MIR evolution.
None of these galaxies had a MIR evolution similar to the variable ECLEs or CLs that had disappeared completely.
We used these six galaxies in our rate calculations as a systematic error to account for the small variations on our measurement of line strengths.
The details and SDSS and DESI spectra of these objects are shown in appendix \ref{app:systematic_cles}.
}

\section{ECLE Rate Analysis}
\label{sec:rate_analysis}
The rate at which a certain type of transient occurs is calculated by dividing the number of transients observed in a given survey, $N$, by the total time that the transients would have been visible in that sample, $t_\mathrm{v}$, i.e.,
\begin{equation}
	R=\frac{N}{\sum t_\mathrm{v}}.
	\label{eq:rate}
\end{equation}
For TDE rates, this is typically normalized by galaxy, so that the rate is given in units of $\mathrm{galaxy}^{-1}~\mathrm{yr}^{-1}$.
We follow this methodology but also calculate mass- and volume-normalized rates.

In our search of SDSS Legacy DR17, we detected \change{16} ECLEs with sufficiently strong CLs.
However, follow-up observations and MIR analysis showed that only \change{five} of the ECLEs might be the result of TDE activity (\citealt{Yang2013,Clark2024}; this work).
\change{As we are unsure of the variable nature of the CLs in SDSS J2220, we did not include it in our rate calculations but added a systematic uncertainty of $+1$ on the number of variable ECLEs detected.
We also include a further $+6$ on this number for the galaxies with moderately strong CLs that are possibly variable.
This systematic accounts for the CL-\shortfspecline{O}{iii} ratio cut used in this work.}

In this section, we first calculate the visibility time of our survey (\secref{subsec:vis_time}).
We then calculate a galaxy-normalized ECLE rate (\secref{subsec:gal_rate}) and a mass-normalized rate, where the visibility time of each galaxy in our sample is weighted by its stellar mass (\secref{subsec:mass_rate}).
Next, we convert the mass-normalized rate to a volumetric rate (\secref{subsec:vol_rate}).
Finally, we compare the ECLE rates measured here to TDE rates from the literature (\secref{subsec:rate_comp}).

\subsection{Visibility time}
\label{subsec:vis_time}

The foundational step of calculating the ECLE rates is determining the length of time that the CLs of an ECLE could be detected at the redshift of each galaxy and QSO in the SDSS Legacy Survey.
This was done by modelling how the CLs of an ECLE evolve over time and combining it with the detection efficiency curve constructed in \secref{subsec:det_eff}.
Below, we describe the process of constructing an ECLE CL strength curve.

For each galaxy in our sample, we determined a peak CL strength and evolved it according to the equation 
\begin{equation}
	S=\alpha t^{\beta},
	\label{eq:power}
\end{equation}
where $S$ is the average strength of the CLs, $t$ is the time since the peak CL strength, and $\alpha$ and $\beta$ are constants.
This was motivated by the proposal that TDE light curves follow a $t^{-\frac{5}{3}}$ power law decline matching the fallback rate of the bound debris \citep{Rees1988,Evans1989,Phinney1989} \change{and that the ECLE MIR declines are fit well by power laws with similar indices \citep{Clark2024}.}
As mentioned in \secref{sec:intro}, different power law indices have been proposed, with the shallowest being $-5/12$.
Therefore, when modelling the CL strength evolution, for each galaxy we randomly sampled $\beta$ from the range $-5/12$ to $-5/3$ assuming a flat probability density function.

\begin{figure}
	\includegraphics[width=0.45\textwidth]{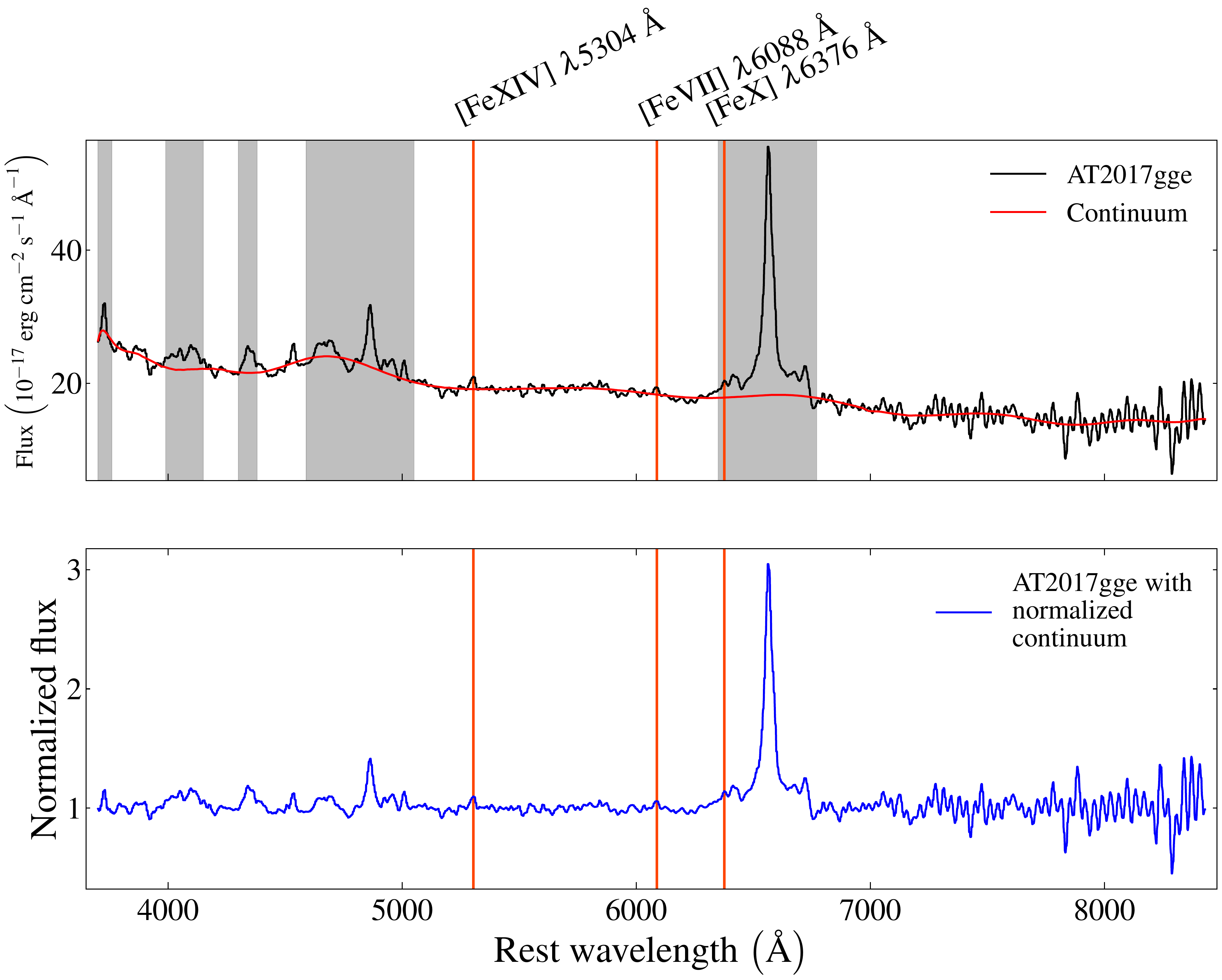}
	\caption{\textit{Top}: Spectrum of confirmed TDE with CLs AT2017gge from 2018 March 9, 218 d post-discovery (black) with continuum fitted using ppxf (red).
	The sections excluded from the fitting procedure are shaded in grey.
	Orange vertical lines mark the CLs visible in the spectrum.
	\textit{Bottom}: Spectrum of AT2017gge with normalized continuum in blue.
	Orange vertical lines mark the CLs visible in the spectrum.
	As the spectrum around the location of the \fspectralline{Fe}{xi}{7894} line is noisy, we are unable to ascertain whether this line does or does not appear in this spectrum.}
	\label{fig:at2017gge_norm}
\end{figure}

\begin{figure}
	\centering
	\includegraphics[width=0.40\textwidth]{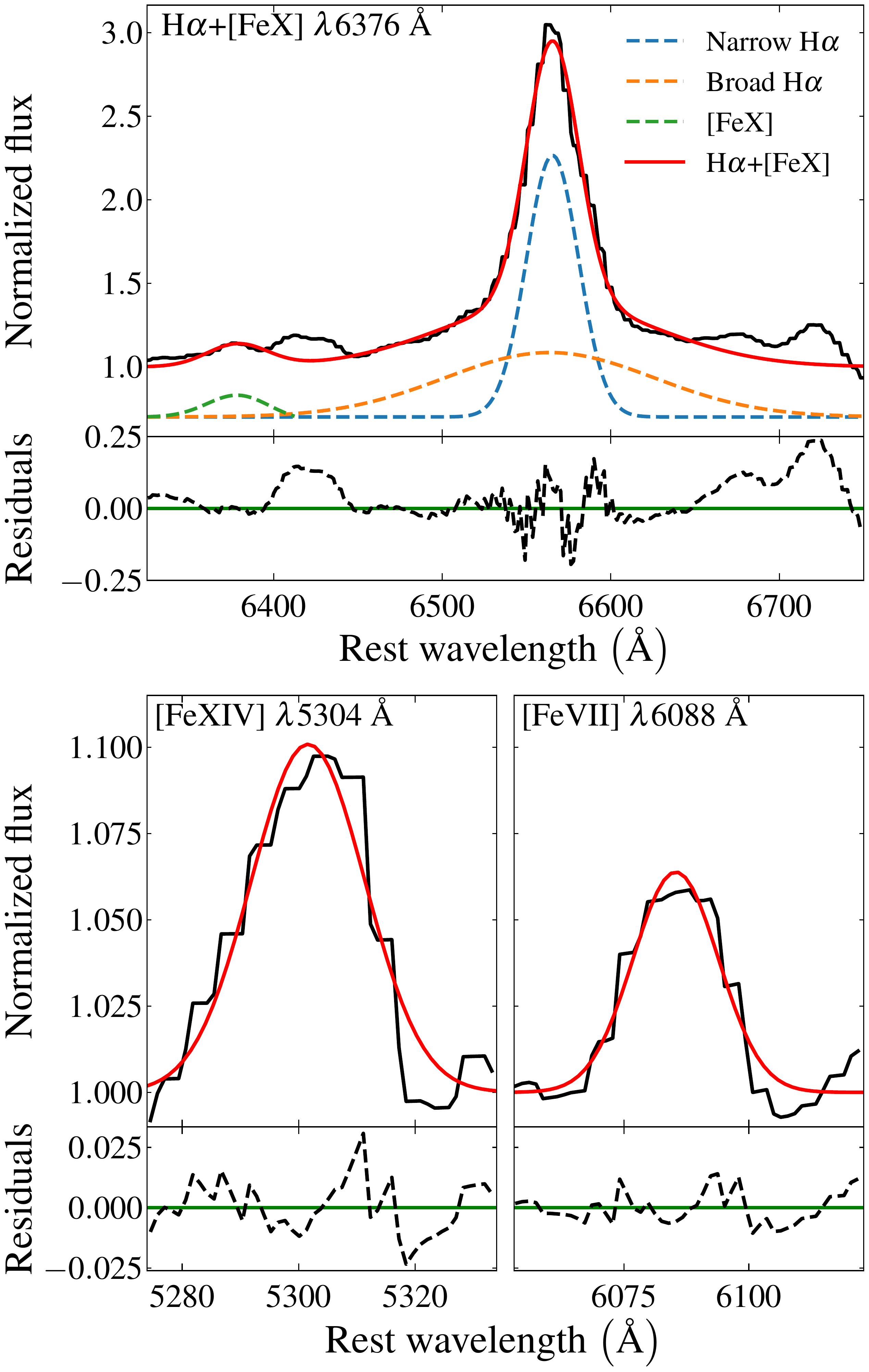}
	\caption{Gaussian fits of the CLs and \Ha complex of the normalized spectrum of AT2017gge and their residuals.
	In all the panels, the normalized spectrum is shown in black and the fitted models in red.
	\textit{Top}: The combined model of the \Ha complex and [\ion{Fe}{X}] line.
	The component Gaussian fits are shown by the offset dashed lines, with the narrow \Ha, broad \Ha, and [\ion{Fe}{X}] fits in blue, orange, and green, respectively.
	\textit{Bottom}: Gaussian fits of [\ion{Fe}{XIV}] and \fspectralline{Fe}{vii}{6088} on the left and right, respectively.}
	\label{fig:CL_fits}
\end{figure}

The bases for the peak CL strengths were determined by using spectroscopic observations of AT2017gge \citep{Onori2022}.
This object was initially discovered as an optical/UV TDE in which there was delayed X-ray emission around 200 d post-discovery, followed by the emergence of [\ion{Fe}{vii}], [\ion{Fe}{X}], and [\ion{Fe}{XIV}] CLs.
These CLs had faded but were still observable 1698 d post-discovery.
\change{This is one of a handful of TDEs confirmed to exhibit CLs in their evolution (see \secref{sec:intro}).
We only use AT2017gge in our rate calculations, as its evolution has been observed for the full phase range that the CLs were present in the \citetalias{Wang2012} sample.}
The first spectrum of AT2017gge that showed CLs was taken on 2018 Mar 9, 218 d post-discovery.
The CLs were weaker in all further spectra, so we took the CLs to be at peak in this first spectrum.
It is important to note that \citet{Onori2022} do not report a detection of \fspectralline{Fe}{xi}{7894} in AT2017gge.
The spectra are consistently noisy in this area, and we are unable to determine if [\ion{Fe}{xi}] is present or not.
Given that all the ECLEs in the \citetalias{Wang2012} sample that exhibit [\ion{Fe}{XIV}] also exhibit [\ion{Fe}{xi}], we consider it unlikely that AT2017gge did not exhibit [\ion{Fe}{xi}] as well.
Therefore, when calculating the average CL strength in AT2017gge, we do not consider the strength of \fspectralline{Fe}{xi}{7894} to be 0 and only take the average of the three CLs reported as present by \citet{Onori2022}.

We measured the strengths of the CLs in AT2017gge by fitting Gaussian curves to the lines and measuring the EW of the curves.
We first normalized the spectrum by fitting the continuum of AT2017gge using the penalized pixel fitting (ppxf) method \citep{Cappellari2023}.
The SDSS spectrum of the host galaxy (SDSS J162034.99+240726.5) was used as a template, which was then fitted to the transient spectrum.
During the fitting process, we excluded the regions around the most extreme emission lines, such as the Balmer emission lines and \spectralline{He}{ii}{4686}.
The spectrum, fitted continuum, and resulting normalized spectrum are shown in \figref{fig:at2017gge_norm}.

We modelled the CLs of AT2017gge with Gaussian functions which were fitted using the \verb|python| package \verb|curve_fit|.
The \fspectralline{Fe}{vii}{6088} and [\ion{Fe}{xiv}] lines were modelled as single Gaussians.
As the [\ion{Fe}{x}] line was near the broad \Ha profile, we modelled the \Ha fit and subtracted it from the [\ion{Fe}{x}] fit to get the true [\ion{Fe}{x}] profile.
The \Ha region was fitted with both single and double (narrow and broad) Gaussian models and an Akaike Information Criterion (AIC) test was used to gauge the model's fit to the data.
For this spectrum, the single Gaussian model gave an AIC value of 392.8 and the double Gaussian model gave 367.9.
The relative likelihood of the double Gaussian model was > 0.95, so it was chosen as the better fit to the data.
We then measured the EWs of the CLs.
The fits of the CLs and \Ha complex are shown in \figref{fig:CL_fits}.

In order to vary the initial line strengths used in \equref{eq:power}, we used a TDE luminosity function (LF) to sample peak luminosities, which we converted to peak CL strengths using AT2017gge.
\change{As the CLs' ionization potentials correspond to X-ray wavelengths, we use the X-ray LF measured by \citet{Sazonov2021}}, which is parametrized as
\begin{equation}
	\frac{\mathrm{d}\dot{N}}{\mathrm{d}\log_{10}L}=\dot{N}_0(L/L_0)^a,
	\label{eq:lf}
\end{equation}
where $\dot{N}$ is the volumetric rate and $L$ is the peak luminosity of the TDE.
\change{The constants have values of $L_0=10^{43}\ \mathrm{erg\ s^{-1}}$, $\dot{N}_0=(1.4\pm0.8)\times10^{-7}\ \mathrm{Mpc^{-3}~yr^{-1}}$ and $a=-0.6\pm0.2$.
As the LF gives the rate of TDEs of a particular peak luminosity, we normalized it over the range of luminosities $10^{42.7}-10^{44.9}\ \mathrm{erg\ s^{-1}}$ to create a probability distribution.}

For each galaxy in the SDSS Legacy Survey, we sampled a peak luminosity, $L_\mathrm{max}$, and converted it into a peak CL strength, $S_\mathrm{max}$, by requiring that the ratio between the peak CL strength and AT2017gge's peak CL strength, $S_\mathrm{gge}$, at 218 d post-discovery was the same as the ratio of the selected peak luminosity and AT2017gge's peak \change{X-ray} luminosity, $L_\mathrm{gge}$, i.e.,
\begin{equation}
	\frac{S_\mathrm{max}}{S_\mathrm{gge}}=\frac{L_\mathrm{max}}{L_\mathrm{gge}}.
	\label{eq:peaks}
\end{equation}
Here we assume that a TDE with a larger peak \change{X-ray} luminosity will produce CLs with a higher peak strength.
$\alpha$ in \equref{eq:power} was then set to equal $S_\mathrm{max}$.

The CL strength evolution was then modelled over a duration of 10 yr.
This period was chosen as all the CLs in the \citetalias{Wang2012} ECLEs had significantly weakened or disappeared 10 yr after they were first detected.
The CL strength curve was then redshifted according to the galaxy's redshift, which increased the time over which the CLs evolved due to cosmological time dilation.
The visibility time, $t_\mathrm{v}$, for each galaxy was then
\begin{equation}
	t_\mathrm{v}=\int\epsilon[S(t)]\mathrm{d}t,
	\label{eq:vis_time}
\end{equation}
where $\epsilon(S)$ is the detection efficiency as a function of CL strength measured in \secref{subsec:det_eff} and the integral runs over the full time over which the strength evolution was modelled.

\subsection{Galaxy-normalized rate}
\label{subsec:gal_rate}
The galaxy-normalized ECLE rate is the number of ECLEs discovered in the SDSS galaxy and QSO sample divided by the sum of the visibility times of all the galaxies searched over,
\begin{equation}
	R_\mathrm{G}=\frac{N_{\mathrm{ECLE}}}{\sum_{i=1}^{N_\mathrm{g}}t_{\mathrm{v},i}},
	\label{eq:gal_rate}
\end{equation}
where $N_{\mathrm{ECLE}}$ is the number of ECLEs detected, $N_\mathrm{g}$ is the number of galaxies searched over, and $t_{\mathrm{v},i}$ is the visibility time of the \textit{i}-th galaxy.

The dominant source of statistical error in our rate calculations is the Poisson uncertainty on the small number of ECLEs detected \citep{Gehrels1986}.
This gives an error on the number of ECLEs detected of \change{$N_\mathrm{ECLE}=5.0~^{+3.4}_{-2.2}$}.
The other sources of statistical uncertainty are the parameters used to calculate the visibility time.
These parameters are the peak average EW of CLs in AT2017gge, the peak \change{X-ray} luminosity of AT2017gge, and the range of power law indices sampled from to construct the CL strength evolutions.
In order to determine the propagation of these uncertainties, we ran a Monte Carlo simulation, in which we calculated the galaxy-normalized rate 500 times with the parameters used in its calculation randomly drawn from their probability density functions each time.
The number of ECLEs was sampled from a Poisson distribution with \change{$\uplambda=5$}.
For the peak AT2017gge CL average EW and \change{X-ray} luminosity used in calculating the peak CL EWs, the parameters were drawn from normal distributions with means set to the measured values and standard deviations set equal to the 1$\upsigma$ errors on the values to account for the uncertainty in the measurement of these parameters.
To account for the uncertainty in the power law indices, we varied the range from which the indices were sampled using an extended range from \citet{Auchettl2017}.
 This analysis found that the range of observed power law indices in TDE light curves is larger than theoretical predictions.
They measured the X-ray power law indices for a sample of 13 TDEs and found that the power law indices ranged from $-0.26\pm0.10$ to $-1.89\pm0.20$.
For the higher (lower) end of the range, we randomly selected a value between the theoretical index, $-5/12$ ($-5/3$), and the highest (lowest) observed index, $-0.26$ ($-1.89$).
Then, for each galaxy in our SDSS DR17 sample, we randomly sampled the power law index from this extended range.
We  turned a histogram of the 500 rates into a probability density function and took the peak of that distribution as the final rate, with $1\upsigma$ statistical errors derived from the distribution.

\change{We also calculated a systematic error on the rate by including SDSS J2220 and the seven galaxies with moderately strong CLs and non-AGN MIR evolutions.}
\mycomment{The systematic errors in our rate calculation come from two sources: the inclusion of SDSS J2220 in our variable ECLE sample and our choice of TDE LF.
We determined the effect these choices had on our final rate by repeating the Monte Carlo simulation using different assumptions.
The first was to remove SDSS J2220 from our variable ECLE sample, leaving the five known variable ECLEs.
During the rate calculation, we sampled the number of ECLEs from a Poisson distribution with $\uplambda=5$.
To determine what effect our choice of TDE LF had, we used a different TDE LF from \citet{Lin2022}.
This LF is parametrized in the same way as \equref{eq:vv_lf} but with $\dot{N}_0=(1.9\pm0.6)\times10^{-8}\ \mathrm{Mpc^{-3}yr^{-1}}$ and $a=-1.22\pm0.18$.
These parameters were fit to the blackbody luminosities of a sample of TDEs from ZTF.}
The rate was recalculated considering \change{these sources} of systematic error using the Monte Carlo simulation. The difference between the original rate and this new rate was taken as the systematic error on the rate.
This process yielded a final galaxy-normalized rate of \galrate.

\begin{figure}
	\centering
	\includegraphics[width=0.45\textwidth]{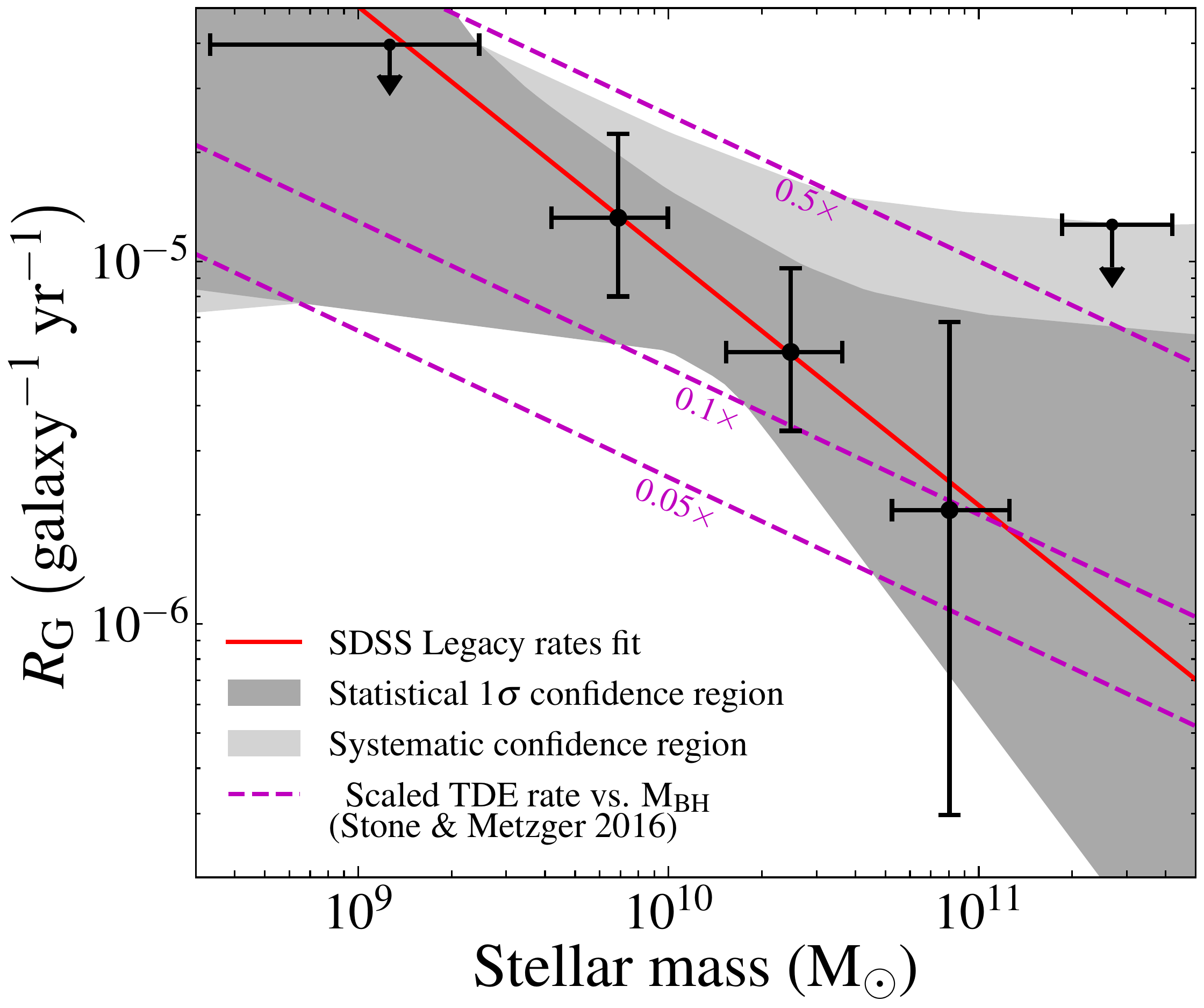}
	\caption{\change{Galaxy-normalized ECLE rates as a function of galaxy stellar mass.
	Vertical error bars show the statistical errors on the rates.
	The horizontal error bars denote the range within each mass bin that includes 68 per cent of the galaxies in that bin.
	The points marked with downward arrows are $2\upsigma$ upper bounds on the rates calculated using the upper Poisson error on zero detections.
	The red curve shows the power law fit to this trend.
	The dark grey region shows the $1\upsigma$ statistical confidence region of the fit, whereas the light grey region shows the additional systematic uncertainty of including SDSS J2220 and the six galaxies with moderately strong CLs.
	The dashed purple lines show the TDE rate vs. BH mass relation from \citet{Stone2016}, scaled by 0.05, 0.1, and 0.5.
	}}
	\label{fig:gal_rate_mass_rel}
\end{figure}

\change{We also investigated whether there was any relation between the galaxies' stellar masses and the galaxy-normalized rate.
This was done by splitting the SDSS sample into mass bins, three of which were equally spaced and contained the variable ECLEs.
The limits of the bins were chosen such that two of the bins contained two of the variable ECLEs each, and the third contained the fifth.
For the low and high mass ends of the SDSS sample where zero variable ECLEs were found, we took the $2\upsigma$ Poisson error on zero detections as upper limits.
The galaxy-normalized rates of each bin were calculated using \equref{eq:gal_rate} and are presented in \figref{fig:gal_rate_mass_rel}. The stellar masses and rates of the SDSS Legacy bins are shown in \tabref{tab:rate_mass_relations}. 
This indicates that ECLE rates are higher in lower-mass galaxies, which aligns with theory predicting that TDEs (and therefore the CLs they produce) will only be visible if they occur around black holes with masses $\lesssim 10^8\ \mathrm{M_\odot}$, which are hosted by galaxies with stellar masses $\lesssim 10^{11}\ \mathrm{M_\odot}$ \citep{Reines2015}.
A similar drop in TDE rates at higher masses is seen in the TDE host galaxy stellar mass function from \citet{vanVelzen2018}.
We fit the trend in \figref{fig:gal_rate_mass_rel} to a power law of the form
\begin{equation}
	\log_{10}(R_\mathrm{G})=a\log_{10}(M)+b,
	\label{eq:rate_rel}
\end{equation}
using the three bins with detected variable ECLEs and requiring that the resultant fit fall below the two upper limits at the SDSS mass extremes.
This fit gave values of $a=-0.7~^{+0.4}_{-0.3},\ b=1.9~^{+3.3}_{-4.4}$ with a reduced $\upchi^2=0.01$.
By including our systematic uncertainties from SDSS J2220 and the six galaxies with moderately strong CLs, this fit flattens slightly, with $a=-0.3~^{+0.3}_{-0.2},\ b=-1.9^{+2.3}_{-2.6}$ with a reduced $\upchi^2=5.9$.
The confidence region for this fit is included in \figref{fig:gal_rate_mass_rel} as a systematic error.
To further investigate this relation, we compared it to the theoretical TDE rate vs. black hole mass relation calculated by \citet{Stone2016}.
The dashed purple lines in \figref{fig:gal_rate_mass_rel} show this relation, scaled down by factors of 0.5, 0.1, and 0.05 to allow comparison with the rates measured here.
The BH mass was converted to galactic stellar mass using the relation from \citet{Reines2015}.
Our power law fit has a similar shape to this relation, and the confidence region lies between the 10 and 50 per cent scaling of the theoretical relation.
This suggests that variable ECLEs are a subset of TDEs.
}

\begin{table}
	\centering
	\caption{\changethree{Variable ECLE rates vs. galaxy stellar mass.}}
	\begin{tabular}{|c|c|c|}
		\hline
		Stellar mass & Galaxy-normalized rate & Mass-normalized rate \\
		$\left(10^{10}~\msun\right)$ & $\left(10^{-6}~\mathrm{galaxy}^{-1}~\mathrm{yr}^{-1}\right)$ & $\left(10^{-17}~\mathrm{M_\odot}^{-1}~\mathrm{yr}^{-1}\right)$ \\
		\hline
		$0.13~^{+0.12}_{-0.09}$ & $<40$ & $<2200$ \\ [1mm]
		$0.7~^{+0.3}_{-0.3}$ & $13~^{+9}_{-5}$ & $160~^{+110}_{-60}$ \\ [1mm]
		$2.5~^{+1.2}_{-0.9}$ & $5.6~^{+4.0}_{-2.2}$ & $19~^{+13}_{-7}$ \\ [1mm]
		$8.0~^{+4.5}_{-2.8}$ & $2~^{+5}_{-2}$ & $2~^{+5}_{-2}$ \\ [1mm]
		$26~^{+15}_{-8}$ & $<13$ & $<3.7$ \\
		\hline
		\multicolumn{3}{l}{The rates marked with $<$ represent $2\upsigma$ upper limits.}
	\end{tabular}
	\label{tab:rate_mass_relations}
\end{table}

\subsection{Mass-normalized rate}
\label{subsec:mass_rate}
In order to calculate the mass-normalized ECLE rate, we weighted the visibility time of each galaxy by its stellar mass, $M_\star$, in the rate calculation, i.e.,
\begin{equation}
	R_\mathrm{M}=\frac{N_{\mathrm{ECLE}}}{\sum_{i=1}^{N_\mathrm{g}}t_{\mathrm{v},i}M_{\star,i}}.
	\label{eq:mass_rate}
\end{equation}
The uncertainties on the number of ECLEs and visibility times were the same as when calculating the galaxy-normalized rate.
Additional uncertainty was introduced by the errors on the measured stellar masses.
We repeated the Monte Carlo simulation to determine how the visibility time and mass uncertainties propagated through to the rate uncertainty.
The masses were drawn from normal distributions with the standard deviation set to the 1$\sigma$ error on the mass.
The mass-normalized rate was calculated to be \massrate.

\begin{figure}
	\centering
	\includegraphics[width=0.45\textwidth]{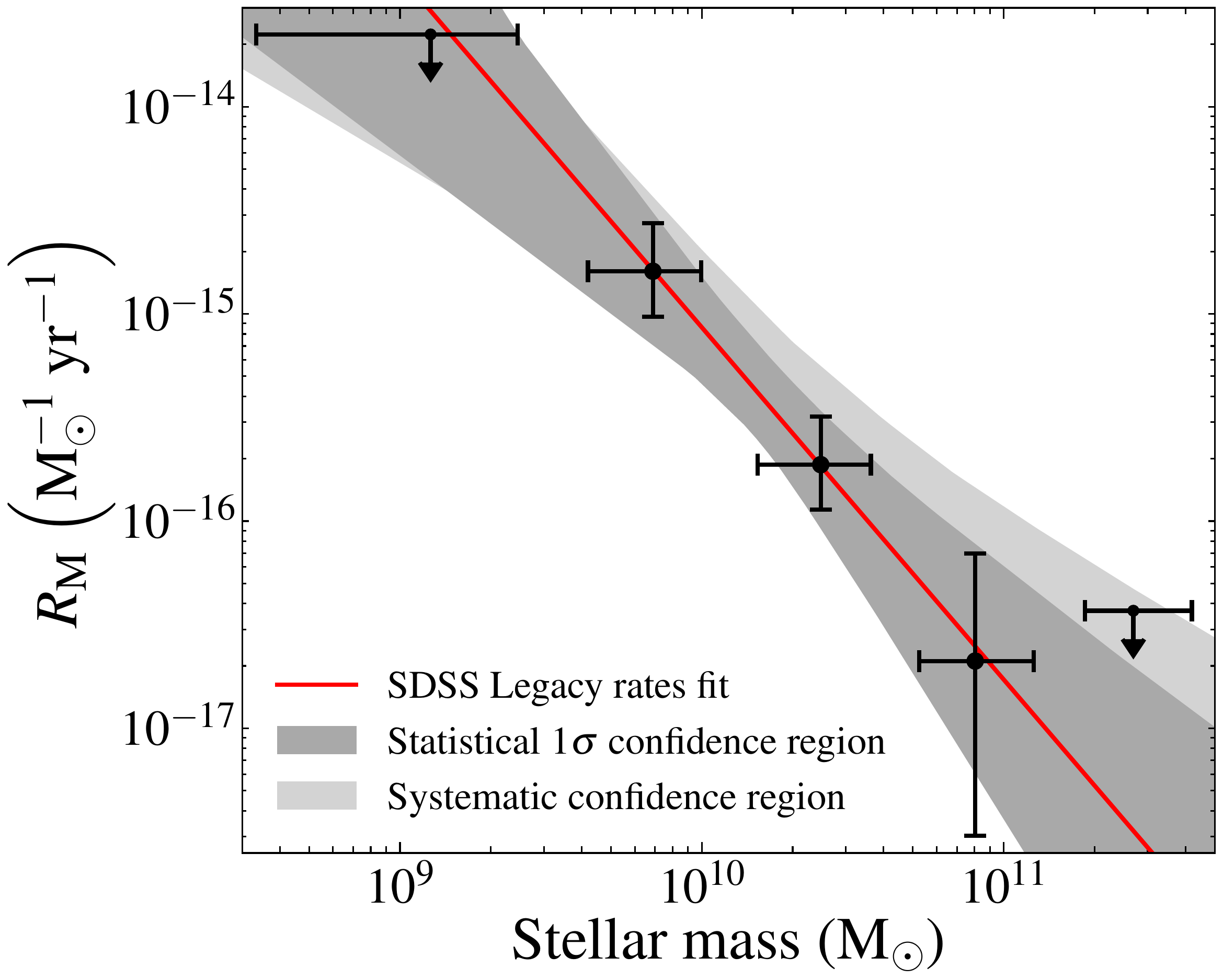}
	\caption{ECLE rates per unit stellar mass as a function of galaxy stellar mass.
	Vertical error bars show the statistical errors on the rates derived using the Monte Carlo simulations detailed above; and the horizontal error bars denote the range within each mass bin that 68 per cent of the galaxies fall.
	The points marked with downward arrows are $2\upsigma$ upper bounds on the rates calculated using the upper Poisson error on zero detections.
	The red line shows the power law fit to this trend.
	The dark grey region shows the $1\upsigma$ confidence on the fit, whereas the light grey shows the systematic on this region from including SDSS J2220 and the six galaxies with moderately strong CLs.
	}
	\label{fig:mass_rate_mass_rel}
\end{figure}

\change{In the same manner as for the galaxy-normalized rate, we split the galaxy sample into mass bins to determine the relation between mass-normalized rate and galactic stellar mass (\figref{fig:mass_rate_mass_rel}).
This reinforces our finding that ECLE rates are higher in lower mass galaxies.
Fitting the three ECLE mass bins to equation \eqref{eq:rate_rel} results in values of $a=-1.7~^{+0.4}_{-0.5},\ b=1.9~^{+4.7}_{-4.5}$ with a reduced $\upchi^2=0.01$.
}

\subsection{Volumetric rate}
\label{subsec:vol_rate}
Following \citet{Graur2013}, we converted our mass-normalized ECLE rate to a volumetric rate by multiplying it by the total cosmic mass density.
This was determined by integrating the galactic stellar mass function (GSMF) measured at $z<0.06$ by \citet{Baldry2012}, $B(M)$, over the mass range of the galaxies in our SDSS sample.
Though the GSMF from \citet{Baldry2012} was constructed only out to $z\sim0.06$, \citet{Hahn2024} have shown that the GSMF does not evolve significantly out to $z\sim0.25$.
Therefore, we are confident in using the \citet{Baldry2012} GSMF across the redshift range of our ECLE sample.
However, the SDSS Legacy Survey has a bias towards higher mass galaxies (see \figref{fig:mass_hist}), which in turn biases our ECLE sample.
As we determined that ECLE rates are higher in lower-mass galaxies, this would serve to lower our volumetric rates.
Therefore, we account for the SDSS Legacy Survey's galaxy mass distribution, weighted by the mass-normalized rate vs. mass relation we found for the sample.
We do this by taking the total cosmic mass density to be the ratio of the integrated GSMF, $B(M)$, to the SDSS Legacy Survey DR17 galaxy-mass distribution, $D(M)$ (which is normalized such that $\int D(M)M\mathrm{d}M=1$), where both mass functions are weighted by the mass-normalized rate vs. mass relation, $R(M)$ found in \secref{subsec:mass_rate}.
The resultant volumetric rate is

\begin{equation}
	R_\mathrm{V}=R_\mathrm{M}\frac{\int B(M)R(M)M\mathrm{d}M}{\int D(M)R(M)M\mathrm{d}M}.
	\label{eq:vol_rate}
\end{equation}

This method added two sources of statistical uncertainty to our final rate value; the errors on the GSMF parameters determined by \citet{Baldry2012} and our fit of the mass-normalized rate vs. mass relation in \secref{subsec:mass_rate}.
We repeated the Monte Carlo simulations to determine how these uncertainties propagated through to the volumetric rate uncertainty.
The parameters were drawn from normal distributions with the standard deviation set to the error on the parameter.
The volumetric rate was calculated to be \volrate.
When not accounting for the SDSS Legacy Survey's mass distribution, the rate was found to be half this value.

The uncertainty budgets of the rates are presented in \tabref{tab:uncertainties}.
The total uncertainty percentages are the linear sum of the total statistical and systematic uncertainties divided by the corresponding rate value.
The uncertainty percentages for each source of uncertainty were calculated by repeating the Monte Carlo simulation, varying the sources of uncertainty independently, and dividing the resulting uncertainty by the corresponding rate.

\changethree{When calculating these uncertainties, we did not consider the properties of the ISM which produces the 
CLs.
Variations in properties such as density, clumpiness, and ionization balance would increase the uncertainties presented here.
}
\begin{table}
	\centering
    \caption{ECLE rate uncertainty budget, in per cent.
	The method for determining the uncertainty percentages is outlined in \secref{subsec:vol_rate}.}
	\begin{tabular}{|l|c|}
		\hline
		Uncertainty 				& Percentage of rate \\
		\hline
		\multicolumn{2}{c}{\textbf{Galaxy-normalized rate}} \\
		Total						& $+212/-49$ \\
		\multicolumn{2}{c}{\textit{Statistical}} \\
		Poisson						& $+67/-43$ \\
		AT2017gge peak CL strength	& $\pm3$ \\
		AT2017gge peak luminosity	& $+50/-15$ \\
		Range of power law indices	& $+19/-12$ \\
		\multicolumn{2}{c}{\textit{Systematic}}	\\		
		SDSS J2220					& $+20/-0$ \\
		CL-\shortfspecline{O}{iii} ratio cut & $+120/-0$ \\
		\hline
		\multicolumn{2}{c}{\textbf{Mass-normalized rate}} \\
		Total						& $+213/-49$ \\
		Galaxy stellar masses 		& $+5/-1$ \\
		\hline
		\multicolumn{2}{c}{\textbf{Volumetric rate}} \\
		Total						& $+411/-64$ \\
		GSMF parameters 			& $+45/-48$ \\
		Rate mass trend fit			& $+348/-32$ \\
		\hline
	\end{tabular}
	\label{tab:uncertainties}
\end{table}

\change{We also investigated the effect our choice of LF to generate starting CL strengths had on our rates.
We recalculated the rates using the optical LF from \citet{vanVelzen2018}, which is parametrized in the same manner as \equref{eq:lf}.
This resulted in rates an order of magnitude higher than those calculated using the X-ray LF.}

\begin{figure*}
	\centering
	\includegraphics[width=\textwidth]{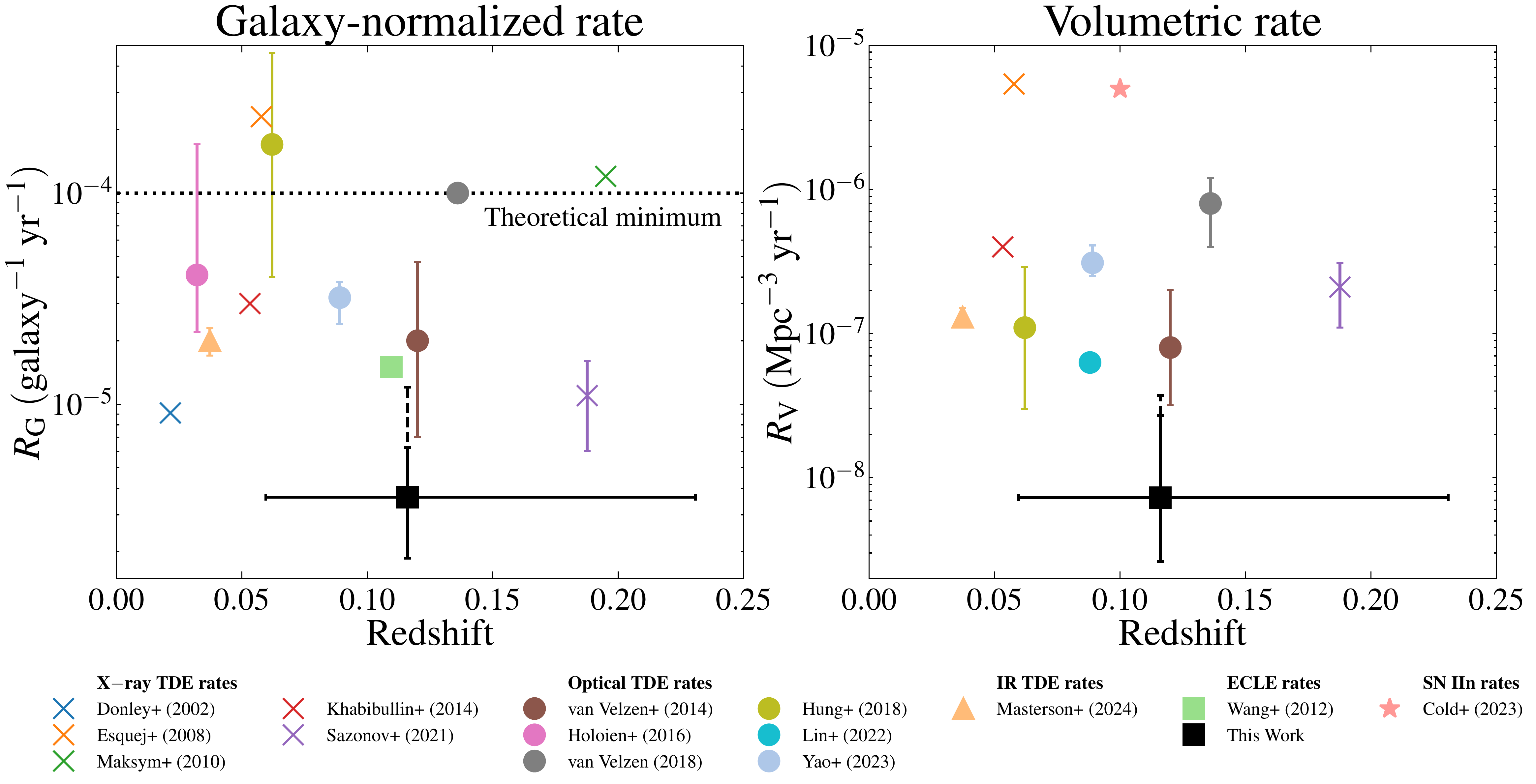}
	\caption{Comparisons of our galaxy-normalized and volumetric ECLE rates with rates from the literature (left and right, respectively).
	TDE rates derived from X-ray surveys are shown as crosses \citep{Donley2002,Esquej2008,Maksym2010,Khabibullin2014,Sazonov2021}, those from optical/UV surveys are shown as circles \citep{vanVelzen2014,Holoien2016,vanVelzen2018,Hung2018,Lin2022,Yao2023}, IR surveys are shown as triangles \citep{Masterson2024}, and ECLE rates are shown as squares \citepalias[this work]{Wang2012}.
	\change{We also include the SN IIn rate from \citet{Cold2023}, shown as a star.}
	Error bars are shown if available.
	The statistical errors on this work are denoted by the solid error bars and the systematic errors by the dashed error bars.
	These errors have been summed linearly, in the same manner as for the uncertainty budget, as described in \secref{subsec:vol_rate}.
	The horizontal error bar shows the range that spans 68 per cent of the SDSS Legacy Survey DR17 sample.
	The dotted horizontal line marks the theoretical minimum TDE rate calculated by \citet{Wang2004}.}
	\label{fig:rate_comp}
\end{figure*}

\subsection{Comparisons to TDE rates}
\label{subsec:rate_comp}
In \figref{fig:rate_comp}, we compare our galaxy-normalized and volumetric ECLE rates to observational rates from the literature \citep{Donley2002,Esquej2008,Maksym2010,Wang2012,Khabibullin2014,vanVelzen2014,Holoien2016,vanVelzen2018,Hung2018,Sazonov2021,Lin2022,Cold2023,Yao2023,Masterson2024}.
\change{Our galaxy-normalized rate is consistent within errors with only three of the observational TDE rates (\citealt{Donley2002}, \citealt{vanVelzen2014}, and \citealt{Sazonov2021}).
It is one to two orders of magnitude lower than the other TDE rates, and nearly two orders of magnitude lower than the theoretical minimum rate from \citet{Wang2004}.
It is also roughly an order of magnitude lower than the ECLE rate estimate from \citetalias{Wang2012}, which was calculated by dividing the number of ECLEs they detected (seven) by the total number of galaxies in their SDSS DR7 spectroscopic sample ($\sim700,000$) and estimating the lifetime of CLs in ECLEs to be three years.
However, this is a rough estimate as \citetalias{Wang2012} did not measure their detection efficiency nor consider that the CLs may not have been variable in some of the ECLEs in their sample.
Our volumetric rate is only consistent within errors with two of the observational TDE rates \citep{vanVelzen2014,Hung2018}, and is one to two orders of magnitude lower than the majority of the other observational rates.}

\change{Both our galaxy-normalized and volumetric ECLE rates are consistent with emission from a subset of TDEs.
This suggests that CLs are only produced in a fraction of TDEs, which likely depends on the nature of the material surrounding the BH.
Study of TDE MIR emission by \citet{Masterson2024} found that the majority of TDE MIR echoes were produced in dusty, star-forming galaxies, which have the necessary material to reprocess the TDE emission to IR wavelengths.
Similarly, we expect that the CLs have been created by material surrounding the BH being ionized by, and reprocessing of the high-energy TDE emission.
This is supported by the fact that all the CL-TDEs discovered so far have occurred in dusty environments, in either star forming or post-starburst galaxies.
\changetwo{\citet{Graur2018} complied a sample of 35 TDE host galaxies, which included 13 star forming galaxies.
If we assume that all TDEs in star forming galaxies produce strong CLs, this would imply $\sim40$ per cent of TDEs can create ECLEs, consistent with our measurements.
However, we note that the \citet{Graur2018} sample is not volume limited and hence does not represent a rate.
}}

\change{By assuming that all the variable ECLEs in our sample are produced by TDEs, we can compare our rates and TDE rates to estimate upper limits on the proportion of TDEs that produce CLs.
Using the $10^{-4}~\mathrm{galaxy}^{-1}~\mathrm{yr}^{-1}$ theoretical minimum galaxy-normalized TDE rate, we estimate that at most $4~^{+3}_{-2}$ per cent of TDEs produce CLs.
Given the discrepancy between theoretical and observational TDE rates and recent work investigating the processes that could lower this theoretical minimum rate \citep{Teboul2023}, this can be considered a conservative lower limit.
Therefore, we also consider the lowest observational galaxy-normalized and volumetric TDE rates, which were measured by \citet{Donley2002} and \citet{Lin2022}, respectively.
From the galaxy-normalized rates, we estimate an upper limit of $40~^{+30}_{-20}$ per cent of TDEs producing CLs. 
The volumetric rates provide a lower value of $12~^{+31}_{-6}$ per cent.
These upper limits derived from observational rates agree within their respective uncertainties, but are both larger than the estimate from the theoretical minimum.
We also note that the range spanned by the estimates from observational rates ($\sim10-40$ per cent) is roughly the same as the range covered by the uncertainty region between the 10 and 50 per cent scalings of the \citet{Stone2016} TDE rate vs. black hole mass relation shown in \figref{fig:gal_rate_mass_rel}.
}

\change{In \figref{fig:rate_comp}, we also include the SN IIn volumetric rate measured by \citet{Cold2023}, as CLs have been observed in SN IIn spectra.
This rate is at least an order of magnitude larger than most of the observational TDE rates and nearly three orders of magnitude larger than our ECLE rate.
Therefore, if SNe IIn are the progenitors of ECLEs then we would expect $\sim0.1$ per cent of them to produce CLs.
Given the large discrepancy between the rates combined with observed CLs from SNe IIn being 100-1000 times weaker than in ECLEs, we consider it very unlikely that SNe IIn are the progenitors of the variable ECLEs studied here.}

\change{A caveat to the findings in this section is that given the small size of our variable ECLE sample, the resulting rates and proportion of CL-TDEs estimates could be largely influenced by small number statistics.
Larger samples of variable ECLEs from spectroscopic galaxy surveys such as DESI will allow us to further constrain these results.
}

\section{Conclusions}
\label{sec:conclusions}
\change{We performed the first full variable ECLE rate calculation, for which we re-searched the SDSS DR17 Legacy Survey for ECLEs.}
We used follow-up spectra and optical and MIR observations to determine the variable nature of the ECLEs discovered.
We then calculated the \change{variable} ECLE rate at a median redshift of 0.1 and compared our results to TDE rates from the literature.
Our conclusions are summarized below.

\begin{itemize}
	\item We discovered \change{nine} new ECLEs among 848,585  galaxies and QSOs from the SDSS DR17 Legacy Survey with sufficiently strong CLs compared to \fspectralline{O}{III}{5007}.
	This \change{more than} doubles the previous sample collected by \citetalias{Wang2012} from SDSS DR7.
	\item Follow-up spectra, MIR \textit{WISE} data, and optical LT data revealed that \change{eight of the nine} new ECLEs were consistent with AGN activity.
	\item The \change{ninth} ECLE, SDSS J2220, varied over time.
	While its behaviour was not identical to that of the five variable ECLEs from the \citetalias{Wang2012} sample, its CLs did fade over time.
	\change{However, its MIR emission was subject to contamination, meaning we could not draw substantial conclusions regarding its nature.
	We did not classify it as a variable ECLE but included it as a systematic in the rate calculations.
	Further follow-up of SDSS J2220 is required to fully establish its variable nature.}
	\item With a sample of \change{five} variable ECLEs, we derived a galaxy-normalized \change{variable} ECLE rate of \galrate.
	We also measured a mass-normalized rate of \massrate\  and converted it to a volumetric rate of \volrate.
	\item \change{\changetwo{Our variable ECLE rates are lower than both theoretical and observational TDE rates by roughly an order of magnitude.
	This suggests that variable ECLEs are due to emission from a subset of roughly 10-40 per cent of all TDEs.}}
	\item This work also reinforces the conclusion reached by \citet{Clark2024} that multiwavelength spectroscopic and photometric follow-up is essential to distinguish variable, TDE-related ECLEs from AGN.
\end{itemize}

\section*{Acknowledgements}
We thank Francesca Onori for sharing the spectra of AT2017gge \change{and Nicholas Stone and the anonymous referee for helpful discussions and comments}.

This work was supported by the Science \& Technology Facilities Council [grants ST/S000550/1 and ST/W001225/1]

Funding for the Sloan Digital Sky Survey (SDSS) and SDSS-II has been provided by the Alfred P. Sloan Foundation, the Participating Institutions, the National Science Foundation, the U.S. Department of Energy, the National Aeronautics and Space Administration, the Japanese Monbukagakusho, and the Max Planck Society, and the Higher Education Funding Council for England. The SDSS Website is http://www.sdss.org/.

The SDSS is managed by the Astrophysical Research Consortium (ARC) for the Participating Institutions. The Participating Institutions are the American Museum of Natural History, Astrophysical Institute Potsdam, University of Basel, University of Cambridge, Case Western Reserve University, The University of Chicago, Drexel University, Fermilab, the Institute for Advanced Study, the Japan Participation Group, The Johns Hopkins University, the Joint Institute for Nuclear Astrophysics, the Kavli Institute for Particle Astrophysics and Cosmology, the Korean Scientist Group, the Chinese Academy of Sciences (LAMOST), Los Alamos National Laboratory, the Max-Planck-Institute for Astronomy (MPIA), the Max-Planck-Institute for Astrophysics (MPA), New Mexico State University, Ohio State University, University of Pittsburgh, University of Portsmouth, Princeton University, the United States Naval Observatory, and the University of Washington.

This material is based upon work supported by the U.S. Department of Energy (DOE), Office of Science, Office of High-Energy Physics, under Contract No. DE-AC02-05CH11231, and by the National Energy Research Scientific Computing Center, a DOE Office of Science User Facility under the same contract. Additional support for DESI was provided by the U.S. National Science Foundation (NSF), Division of Astronomical Sciences under Contract No. AST-0950945 to the NSF's National Optical-Infrared Astronomy Research Laboratory; the Science and Technology Facilities Council of the United Kingdom; the Gordon and Betty Moore Foundation; the Heising-Simons Foundation; the French Alternative Energies and Atomic Energy Commission (CEA); the National Council of Science and Technology of Mexico (CONACYT); the Ministry of Science and Innovation of Spain (MICINN), and by the DESI Member Institutions: \url{https://www.desi.lbl.gov/collaborating-institutions}. Any opinions, findings, and conclusions or recommendations expressed in this material are those of the author(s) and do not necessarily reflect the views of the U. S. National Science Foundation, the U. S. Department of Energy, or any of the listed funding agencies.

The authors are honored to be permitted to conduct scientific research on Iolkam Du'ag (Kitt Peak), a mountain with particular significance to the Tohono O'odham Nation.

This research has made use of NASA's Astrophysics Data System Bibliographic Services and the NASA/IPAC Infrared Science Archive, which is funded by the National Aeronautics and Space Administration (NASA) and operated by the California Institute of Technology. This publication also makes use of data products from NEOWISE, which is a project of the Jet Propulsion Laboratory/California Institute of Technology, funded by the Planetary Science Division of NASA. The CRTS survey is supported by the U.S. National Science Foundation (NSF) under grants AST-0909182 and AST-1313422.

Based on observations obtained at the international Gemini Observatory, a program of NSF's NOIRLab, processed using DRAGONS (Data Reduction for Astronomy from Gemini Observatory North and South), which is managed by the Association of Universities for Research in Astronomy (AURA) under a cooperative agreement with the National Science Foundation on behalf of the Gemini Observatory partnership: the National Science Foundation (United States), National Research Council (Canada), Agencia Nacional de Investigaci\'{o}n y Desarrollo (Chile), Ministerio de Ciencia, Tecnolog\'{i}a e Innovaci\'{o}n (Argentina), Minist\'{e}rio da Ci\^{e}ncia, Tecnologia, Inova\c{c}\~{o}es e Comunica\c{c}\~{o}es (Brazil), and Korea Astronomy and Space Science Institute (Republic of Korea). This work was enabled by observations made from the Gemini North telescope, located within the Maunakea Science Reserve and adjacent to the summit of Maunakea. We are grateful for the privilege of observing the Universe from a place that is unique in both its astronomical quality and its cultural significance.

\section*{Data Availability}
The data underlying this article are available in the article and in its online supplementary material through Zenodo at \citet{Callow2024}. The reduced data and derived measurements in this article will be shared on reasonable request to the corresponding author.

\bibliographystyle{mnras}
\bibliography{sdss_ecles}

\appendix

\section{Observing Conditions}
\label{app:obs_conditions}

Here we summarize the observing conditions during the spectroscopic observations used in this work.

\begin{table}
    \centering
    \caption{Summary of the spectroscopic observing conditions.}
    \begin{tabular}{|c|c|c|c|}
        \hline
        SDSS Short Name & MJD & Seeing (arcsec) & Air mass \\
        \hline
        \multicolumn{4}{c}{\bf{SDSS Legacy}} \\
		SDSS J0807 & 53682 & 1.4 & 1.0 \\
		SDSS J1207 & 54484 & 1.7 & 1.4 \\
        SDSS J1238 & 54481 & 1.6 & 1.0 \\
		SDSS J1247 & 53876 & 1.1 & 1.3 \\
        SDSS J1402 & 54178 & 1.2 & 1.1 \\
        SDSS J1458 & 54539 & 1.6 & 1.0 \\
		SDSS J1459 & 53146 & 2.2 & 1.1 \\
        SDSS J1715 & 51788 & 1.7 & 1.3 \\
        SDSS J2220 & 52203 & 1.6 & 1.4 \\
        \hline
        \multicolumn{4}{c}{\bf{DESI}} \\
        SDSS J0807 & 59646 & 1.0 & 1.2 \\
        SDSS J1238 & 60306 & 1.0 & 1.1 \\
        SDSS J1247 & 59732 & 0.8 & 1.1 \\
        SDSS J1402 & 60058 & 1.4 & 1.0 \\
        SDSS J1458 & 59378 & 0.9 & 1.0 \\
        SDSS J1459 & 59683 & 1.1 & 1.0 \\
        SDSS J2220 & 59846 & 0.9 & 1.3 \\
        \hline
        \multicolumn{4}{c}{\bf{GMOS}} \\
        SDSS J1207 & 60489 & 1.5 & 1.7 \\
        SDSS J1715 & 60154 & 1.3 & 1.5 \\
        SDSS J2220 & 60155 & 0.8 & 1.2 \\
        \hline
    \end{tabular}
    \label{tab:obs_conditions}
\end{table}

\section{ECLE DESI Spectra}
\label{app:desi_specs}

Here we list the TARGETIDs of the DESI spectra of the ECLEs found in this work.
We also include the surveys and programs that these objects were part of.

\begin{table}
    \centering
    \caption{Summary of the DESI spectra of the new ECLEs.}
    \begin{tabular}{|c|c|c|}
        \hline
        SDSS Short Name & DESI TARGETID & Survey \& Program \\
        \hline
		SDSS J0807 & 39628121963496839 & BGS Bright \\
        SDSS J1238 & 39628239013940096 & BGS Bright \\
		SDSS J1247 & 39627956754055903 & QSO        \\
		SDSS J1402 & 39628475925004934 & BGS Bright \\
        SDSS J1458 & 39628245288616742 & LRG        \\
		SDSS J1459 & 39633091307767094 & BGS Bright \\
		SDSS J2220 & 39627597541281704 & BGS Bright \\
        \hline
    \end{tabular}
    \label{tab:desi_specs}
\end{table}

\section{Galaxies with moderately strong CLs}
\label{app:systematic_cles}

Here we summarize the properties of the spectra of the galaxies with moderately strong CLs that were included in the systematic uncertainty calculations on the rates.
We also show the spectra themselves.

\begin{table*}
    \centering
    \caption{Summary of the galaxies with moderately strong CLs, six of which were included as a positive systematic in the rate calculations.}
    \begin{tabular}{|c|c|c|c|c|c|c|}
        \hline
        SDSS Name & RA (J2000) & Dec (J2000) & Redshift & DESI TARGETID & Survey \& Program & Included \\
          &  &  &  &  &  & as systematic \\
        \hline
        SDSS J080523.39+281814.0 & 08:05:23.3970 & +28:18:14.0623 & 0.129 &  &  & \cmark \\
        SDSS J080625.29+160416.3 & 08:06:25.2964 & +16:04:46.3145 & 0.092 & 39628168700626077 & BGS Bright & \xmark \\
        SDSS J083053.67+371802.9 & 08:30:53.6795 & +37:18:02.9202 & 0.099 &  &  & \cmark \\
        SDSS J085737.78+052821.3 & 08:57:37.7805 & +05:28:21.3357 & 0.059 & 39627919739324361 & BGS Bright & \xmark \\
        SDSS J095434.81+501433.3 & 09:54:34.8184 & +50:14:33.3330 & 0.053 & 39633251614066475 & BGS Bright & \xmark \\
        SDSS J110649.90+101738.4 & 11:06:49.8013 & +10:17:38.4346 & 0.168 & 39628034029914999 & BGS Bright & \xmark \\
        SDSS J114114.63+055952.8 & 11:41:14.6395 & +05:59:52.8549 & 0.099 & 2305843038338946672 & BGS Bright & \xmark \\
        SDSS J115710.68+221746.2 & 11:57:10.6829 & +22:17:46.2312 & 0.052 &  &  & \cmark \\
        SDSS J121754.97+583935.6 & 12:17:54.9783 & +58:39:35.6511 & 0.023 &  &  & \cmark \\
        SDSS J123235.82+060309.9 & 12:32:35.8213 & +06:03:09.9785 & 0.083 & 39627932666169508 & QSO & \xmark \\
        SDSS J134234.30+191338.9 & 13:42:34.3096 & +19:13:38.9481 & 0.086 & 39628244986626198 & BGS Bright & \cmark \\
        SDSS J150107.62+160741.4 & 15:01:07.6261 & +16:07:41.4843 & 0.190 &	39628176183263704 & BGS Bright & \cmark \\
        SDSS J161809.36+361957.8 & 16:18:09.3692 & +36:19:57.8907 &	0.034 &	39633006326975257 & BGS Bright & \xmark \\
        SDSS J163631.28+420242.5 & 16:36:31.2890 & +42:02:42.5536 & 0.061 & 39633114334495784 & BGS Bright & \xmark \\
        \hline
    \end{tabular}
    \label{tab:sys_sample}
\end{table*}

\begin{figure*}
    \centering
    \includegraphics[width=\textwidth]{images/weak_CLEs_SDSS_vs_DESI_Comparison_Mosaic_0.pdf}
    \caption{Comparisons of the spectra of the galaxies with moderately strong CLs from SDSS (black) and DESI (red).
    Emission lines of interest are marked by vertical lines.
    The dotted lines in the upper plots indicate the region used to rescale the spectra with respect to each other to allow for easy comparison.
    For the line-specific plots, this rescaling was done on continuum sections of the spectra near the emission line.}
    \label{fig:weak_cle_spec_mosaic}
\end{figure*}

\begin{Contfigure*}
    \centering
    \includegraphics[width=\textwidth]{images/weak_CLEs_SDSS_vs_DESI_Comparison_Mosaic_1.pdf}
    \caption{Comparisons of the spectra of the galaxies with moderately strong CLs from SDSS (black) and DESI (red).
    Emission lines of interest are marked by vertical lines.
    The dotted lines in the upper plots indicate the region used to rescale the spectra with respect to each other to allow for easy comparison.
    For the line-specific plots, this rescaling was done on continuum sections of the spectra near the emission line.}
\end{Contfigure*}

\begin{Contfigure*}
    \centering
    \includegraphics[width=\textwidth]{images/weak_CLEs_SDSS_vs_DESI_Comparison_Mosaic_2.pdf}
    \caption{Comparisons of the spectra of the galaxies with moderately strong CLs from SDSS (black) and DESI (red).
    Emission lines of interest are marked by vertical lines.
    The dotted lines in the upper plots indicate the region used to rescale the spectra with respect to each other to allow for easy comparison.
    For the line-specific plots, this rescaling was done on continuum sections of the spectra near the emission line.}
\end{Contfigure*}

\begin{Contfigure*}
    \centering
    \includegraphics[width=\textwidth]{images/weak_CLEs_SDSS_vs_DESI_Comparison_Mosaic_3.pdf}
    \caption{Comparisons of the spectra of the galaxies with moderately strong CLs from SDSS (black) and DESI (red).
    Emission lines of interest are marked by vertical lines.
    The dotted lines in the upper plots indicate the region used to rescale the spectra with respect to each other to allow for easy comparison.
    For the line-specific plots, this rescaling was done on continuum sections of the spectra near the emission line.}
\end{Contfigure*}

\begin{Contfigure*}
    \centering
    \includegraphics[width=\textwidth]{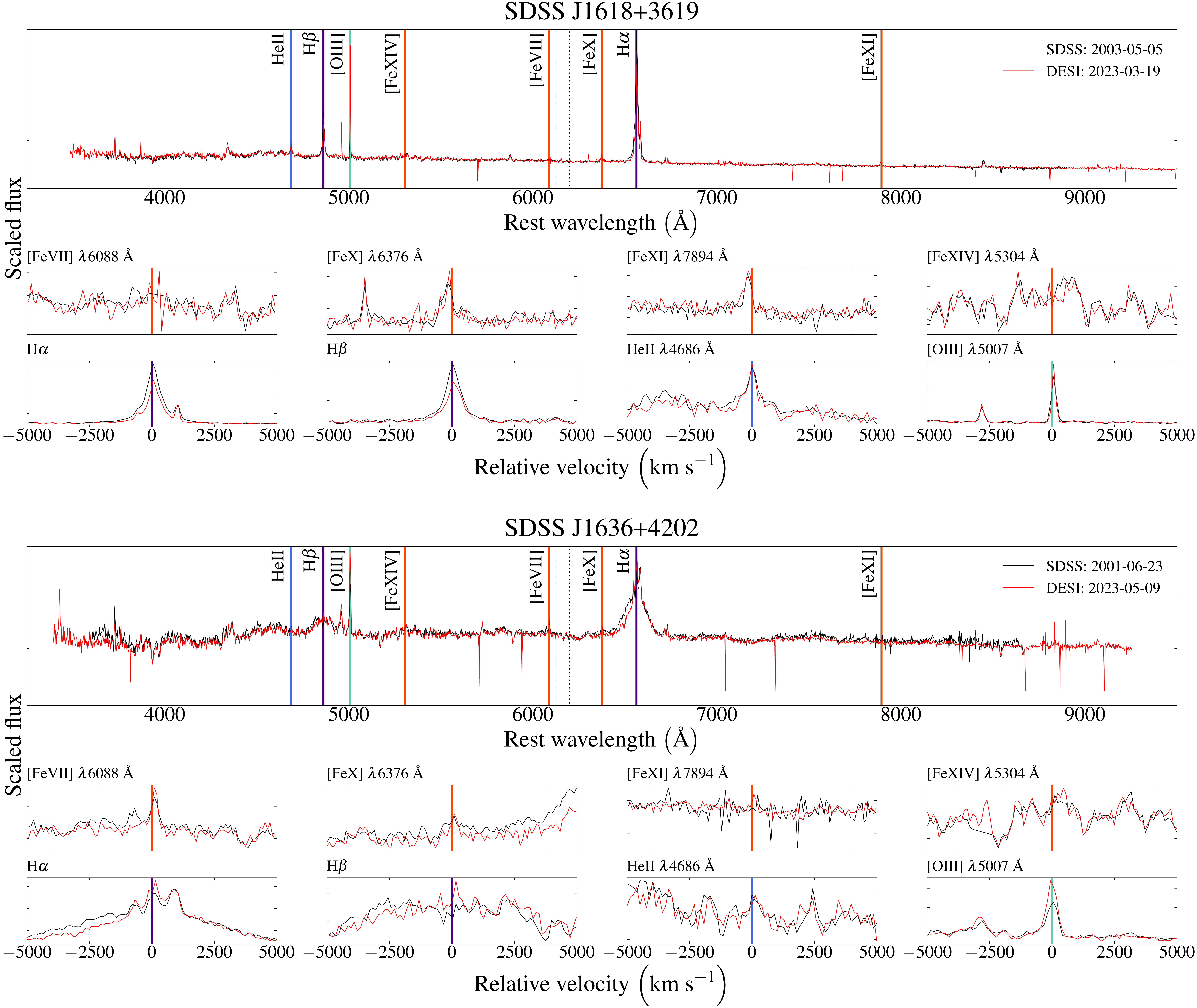}
    \caption{Comparisons of the spectra of the galaxies with moderately strong CLs from SDSS (black) and DESI (red).
    Emission lines of interest are marked by vertical lines.
    The dotted lines in the upper plots indicate the region used to rescale the spectra with respect to each other to allow for easy comparison.
    For the line-specific plots, this rescaling was done on continuum sections of the spectra near the emission line.}
\end{Contfigure*}

\bsp
\label{lastpage}
\end{document}